\documentclass[sigconf]{acmart}

\AtBeginDocument{%
  }

\setcopyright{acmlicensed}
\copyrightyear{2025}
\acmYear{2025}
\acmDOI{XXXXXXX.XXXXXXX}
\acmConference[Conference acronym 'XX]{Make sure to enter the correct
  conference title from your rights confirmation email}{June 03--05,
  2018}{Woodstock, NY}
\acmISBN{XXXX-X/2025/03}

\usepackage{amsmath,amssymb,amsfonts}
\usepackage{xcolor}
\usepackage{algorithmic}
\usepackage{graphicx}
\usepackage{textcomp}
\usepackage{algorithm}
\usepackage{algorithmic}

  \usepackage{rotating}
        \usepackage{multirow}
        \usepackage{lscape}
\usepackage{color,soul}
\usepackage{xcolor}
\usepackage{caption}
\usepackage{subcaption}
\usepackage{makecell}
\usepackage{float}
\usepackage{pifont}
\usepackage{comment}

\begin{document}

\title{ Large Language Models (LLMs) for Source Code Analysis: applications, models and datasets  \\  
}

\author{Hamed Jelodar, Mohammad Meymani, Roozbeh Razavi-Far }
\email{ {h.jelodar, mohammad.meymani79,roozbeh.razavi-far}@unb.ca }

\affiliation{%
  \institution{Canadian Institute for Cybersecurity, Faculty of Computer Science, University of New Brunswick}
  \city{Fredericton}
  \state{NB}
  \country{Canada}
}









\date{Received: date / Accepted: date}

\begin{abstract}
Large language models (LLMs) and transformer-based architectures are increasingly utilized for source code analysis. As software systems grow in complexity, integrating LLMs into code analysis workflows becomes essential for enhancing efficiency, accuracy, and automation. This paper explores the role of LLMs for different code analysis tasks, focusing on three key aspects: \textit{ 1) what they can analyze and their applications, 2) what models are used and 3) what datasets are used, and the challenges they face}. Regarding the goal of this research, we investigate scholarly articles that explore the use of LLMs for source code analysis to uncover research developments, current trends, and the intellectual structure of this emerging field. Additionally, we summarize limitations and highlight essential tools, datasets, and key challenges, which could be valuable for future work.


\end{abstract}
\maketitle
\section{Introduction}
\begin{figure}[H]
    \centering
    \includegraphics[width=\linewidth]{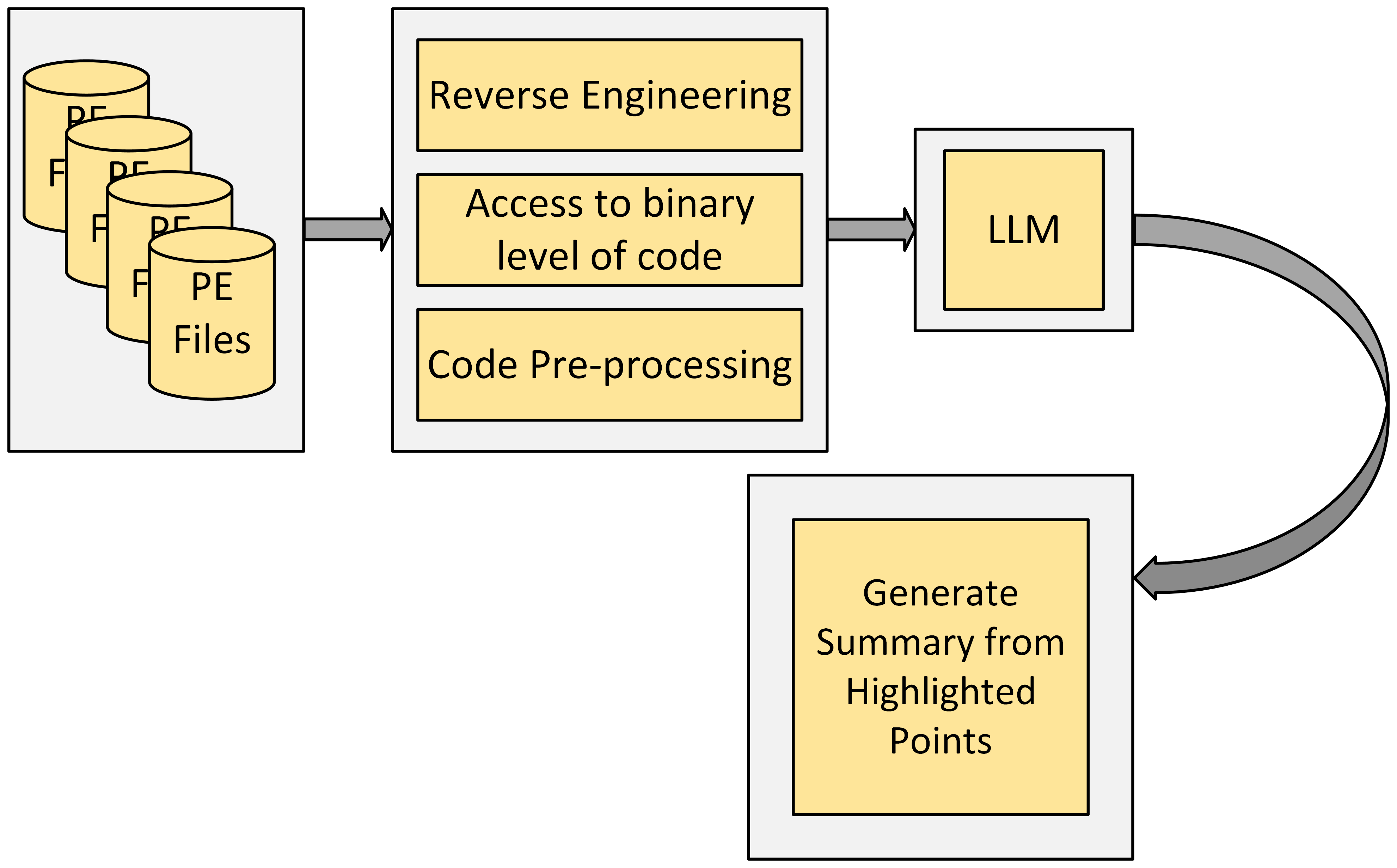}
    \caption{Binary-code summary using LLM}
    \label{fig:Binary-code summary using LLM}
 \end{figure}

In a broader context, large language models (LLMs) have been increasingly applied in the analysis of source code, leading to advancements in software engineering, bug detection, and code optimization \cite{131}. LLMs are particularly useful for understanding code patterns, structure, and functionality. They help identify hidden patterns within large codebases, improve code quality, and automate tasks. By extracting semantic structures from complex source code, LLMs enable developers to better understand the behavior and logic of code, facilitating easier maintenance and enhancement. These models also play a key role in understanding how different pieces of code interact, much like how they are used to analyze interactions. LLMs are applied in various fields for source codes, including code summary \cite{28,29,30,31} \cite{41,47,48}, code generation \cite{50,51,52,53}, showcasing their versatility and potential in both real-world application and research. For instance, in code summarization, the authors in \cite{30} explored LLMs to generate concise natural language descriptions of code snippets. It investigates the effectiveness of different prompting techniques, evaluates various LLMs’ code summarization abilities, and explores how model settings affect performance. The study finds that simpler prompting techniques like zero-shot prompting may outperform more advanced ones and that the impact of LLM settings varies by language. Additionally, LLMs struggle more with logic programming languages. Finally, they found that CodeLlama-Instruct with 7B parameters is better than other methods for code-summary tasks. Fig \ref{fig:Binary-code summary using LLM} show a process of code-summary based on an LLM model. Also, in other works, the authors in \cite{zhang2025unveiling} analyzed biases in LLMs that unintentionally favored service providers like Google or Amazon during code generation. Using a dataset of 17,014 prompts, they found that these models exhibited strong preferences for these providers, even without specific user requests. This highlighted concerns about the biases in AI-generated content, particularly in coding applications.\\

Despite the growing interest in utilizing LLMs for source code analysis, as evidenced by numerous recent papers, the research community still lacks a comprehensive survey that consolidates existing LLM-based code analysis studies, identifies the challenges encountered, and outlines future research directions \cite{197}. We aim to fill this gap by conducting a systematic survey of LLM applications in various source code analysis tasks. \\

Generally, this survey will examine different perspectives, such as foundational theories, technical approaches, datasets, performance benchmarks, and future research opportunities. We believe that this survey will provide a clear overview of the current state of the field and highlight promising avenues for future advancements in this rapidly evolving research area.

Since LLMs have demonstrated exceptional capabilities in natural language processing, they are becoming increasingly important for understanding and analyzing source code. Our motivation is to investigate the application of LLMs in source code analysis, with a focus on various aspects such as models, datasets, and practical applications. The main goal of this work is to provide a comprehensive overview of how LLMs can be utilized for source code analysis. In summary, this paper makes four main contributions:
\begin{itemize}
  \item We investigate scholarly articles that explore the use of LLMs for source code analysis to uncover research developments, current trends, and the intellectual structure of this emerging field
  \item We examine the applications of LLMs in various areas of code-analysis challenges, such as code generation, code decompiling and code summarization.
  \item We review and introduce prominent recent LLM models (Main/fine-tuned models) for code analysis.
  \item We introduce some of the most widely used datasets for LLM-based source code analysis, highlighting their strengths and limitations.  
\end{itemize}

There are a few studies that could be related to our research \cite{1,60,126,167}. For example, in \cite{1}, the authors investigated the role of LLMs in code understanding. In \cite{60,115,200}, only code generating ability of LLMs is addressed. In \cite{126}, code understanding and code generation are addressed. In \cite{167}, code summarization, code generation and security analysis are addressed. To the best of our knowledge, this is the first work to investigate different aspects of language models in code-related tasks, including code understanding, code disassembly, code decompilation, code summarization, code generation, comment generation, and security analysis.\\

Moreover, we provide a novel code-analysis taxonomy, classifying important LLMs according to their respective families as shown in Figure \ref{fig:Models-for-code-taxonomy}. Additionally, we include notable fine-tuned models in our taxonomy, emphasizing their specific enhancements and applications in various code-related tasks. In Figure \ref{fig:LLM-Datasets-Timeline}, we present a comprehensive timeline of free datasets for code evaluation, showcasing their evolution and impact on the field. Furthermore, in Table \ref{tab:comparison-of-our-work-with-related-surveys}, we illustrate the coverage of code-related tasks in our survey compared to related surveys, highlighting key gaps and contributions that differentiate our work.

\begin{table}[H]
    \centering
    \caption{Our work vs. other surveys- Code Understanding (CU), Code Disassembling (CD), Code Decompiling (CdC), Code Summarization (CS), Code Generation (CG), Comment Generation (ComG), Security Analysis (SA)}
    \begin{tabular}{c c c c c c c c}
        \hline
        Survey & \multicolumn{7}{c}{Code-related Tasks} \\
        \hline
         &CU&CD&CdC&CS&CG&ComG&SA\\
         \hline

        \cite{60}&\ding{55}&\ding{55}&\ding{55}&\ding{55}&\checkmark&\ding{55}&\ding{55}\\

        \cite{115}&\ding{55}&\ding{55}&\ding{55}&\ding{55}&\checkmark&\ding{55}&\ding{55}\\

        \cite{126}&\checkmark&\ding{55}&\ding{55}&\ding{55}&\checkmark&\ding{55}&\ding{55}\\

        \cite{1}&\checkmark&\ding{55}&\ding{55}&\ding{55}&\ding{55}&\ding{55}&\ding{55}\\

        \cite{167}&\ding{55}&\ding{55}&\ding{55}&\checkmark&\checkmark&\ding{55}&\checkmark\\

        \cite{200}&\ding{55}&\ding{55}&\ding{55}&\ding{55}&\checkmark&\ding{55}&\ding{55}\\
         
         Our Work & \checkmark & \checkmark & \checkmark & \checkmark & \checkmark & \checkmark & \checkmark\\
         \hline
    \end{tabular}
    \label{tab:comparison-of-our-work-with-related-surveys}
\end{table}

The rest of the paper is organized as follows. In section \ref{sec:background}, we provide a background on traditional tools, non-LLM Models, and LLM models for source code analysis. In section \ref{sec:applications-of-LLM-for-source-code-analysis} we discuss a wide range of applications for source code analysis, providing a comprehensive table for each application. In section \ref{sec:Pre-training for Domain Adaptation on code analysis}, we discuss the area of domain adaptation on code analysis. In section \ref{sec:Popular Models: Language models for Source
Code Analysis}, we discuss the most popular models for source code analysis, providing a wide, novel taxonomy for these models. In section \ref{sec:Datasets and LLM code analysis}, we discuss free benchmarks for code analysis, providing a table and a timeline for these datasets. In section \ref{sec:Discussion and Future works}, we address the potential future gaps, discussing the current limitations of the existing models and providing potential solutions. Finally, in Section \ref{sec:Conclusion}, we provide a summary of our findings.

\section{BACKGROUND}
\label{sec:background}
This section explains NLP and LLM techniques for source code analysis, providing a background on how these methods can be applied to source code analysis.

\subsection{ Traditional tools for source  code analysis} 
There are traditional tools that can be helpful for source code analysis. For real-time debugging and analysis, dynamic analysis tools and debuggers like OllyDbg\cite{133} and x64dbg \cite{132} enable analysts to inspect the behavior of running programs.

These tools provide the ability to interact with a program during execution, enabling users to modify its behavior, observe runtime data, and identify vulnerabilities or bugs.

\begin{figure*}[!htbp]
\centering
\includegraphics[width=\linewidth]{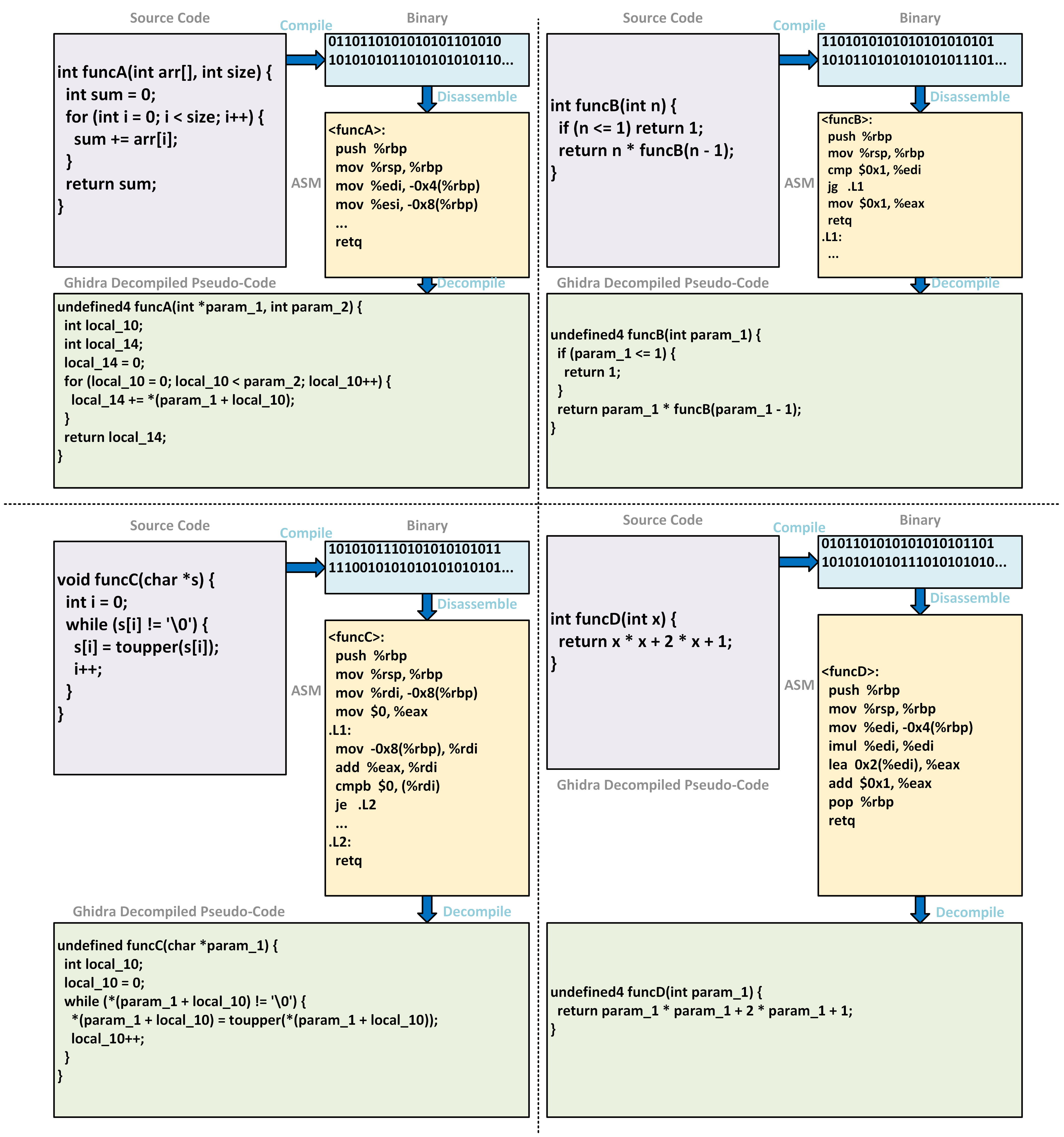}
\caption{The Process of Code Decompiling in Different tasks using Ghidra.}
\label{fig:x cubed graph}
\end{figure*}
Decompiling and disassembling tools \cite{19,22} are essential for reverse engineering compiled binaries, allowing analysts to inspect and understand the internal workings of a program. Disassemblers such as Ghidra \cite{134} are widely used for converting binary code into assembly language, serving as a non-LLM method, as illustrated in Fig 2.

These tools break down machine code into human-readable assembly instructions, helping reverse engineers analyze and debug low-level program behavior. Disassemblers are especially valuable for understanding how a program operates at the CPU instruction level and are frequently used in malware analysis and security research \cite{23}. Capstone \cite{136} and Binary Ninja \cite{135} are also key players in this category, providing lightweight disassembly capabilities for a range of CPU architectures .

\subsection{Non-LLM models for code analysis}
Traditional NLP methods for source code analysis focus on taking advantage of the structured nature of programming languages using statistical \cite{138}, rule-based \cite{137}, and classical machine learning techniques. One of the foundational steps is lexical analysis\cite{139}, where code is tokenized into meaningful units like keywords, operators, and identifiers. This process enables tasks like syntax highlighting, code formatting, and even plagiarism detection. In parallel, syntax analysis uses parsers to generate Abstract Syntax Trees (ASTs)\cite{120}, capturing the hierarchical structure of the code for applications such as syntax error detection and automated code completion \cite{120,121,122,123} . \\

Moreover, semantic analysis examines the deeper meaning of the code, including type checking and symbol resolution \cite{140}. Complementing this, feature engineering techniques manually extract metrics such as cyclomatic complexity, coupling, and dependencies to assess code quality and maintainability. Another important area is code similarity \cite{141,142} and clone detection \cite{143,144}, which employs techniques like token-based comparisons and AST matching to identify duplicate code segments.  Additionally, information retrieval techniques like cosine similarity \cite{145,146} enable effective code search and plagiarism detection, making them valuable for large-scale code repositories.

 \subsection{Transformer and LLM Models for Code}  
Transformers and LLMs have revolutionized source code analysis and generation, using their ability to process both natural and programming languages effectively \cite{115,124,126,127,129}. Fig. 3 shows a general view of the application of LLMs for source code analysis based on different NLP techniques. In more detail, some researchers have used pre-trained techniques or LLM models for code analysis. For example, OpenAI’s Codex \cite{106} and GPT-4, along with specialized models such as CodeBERT \cite{114} and GraphCodeBERT \cite{147}, are trained on large code datasets and related documentation.\\

\begin{figure}[h]
\centering
\includegraphics[width=\linewidth]{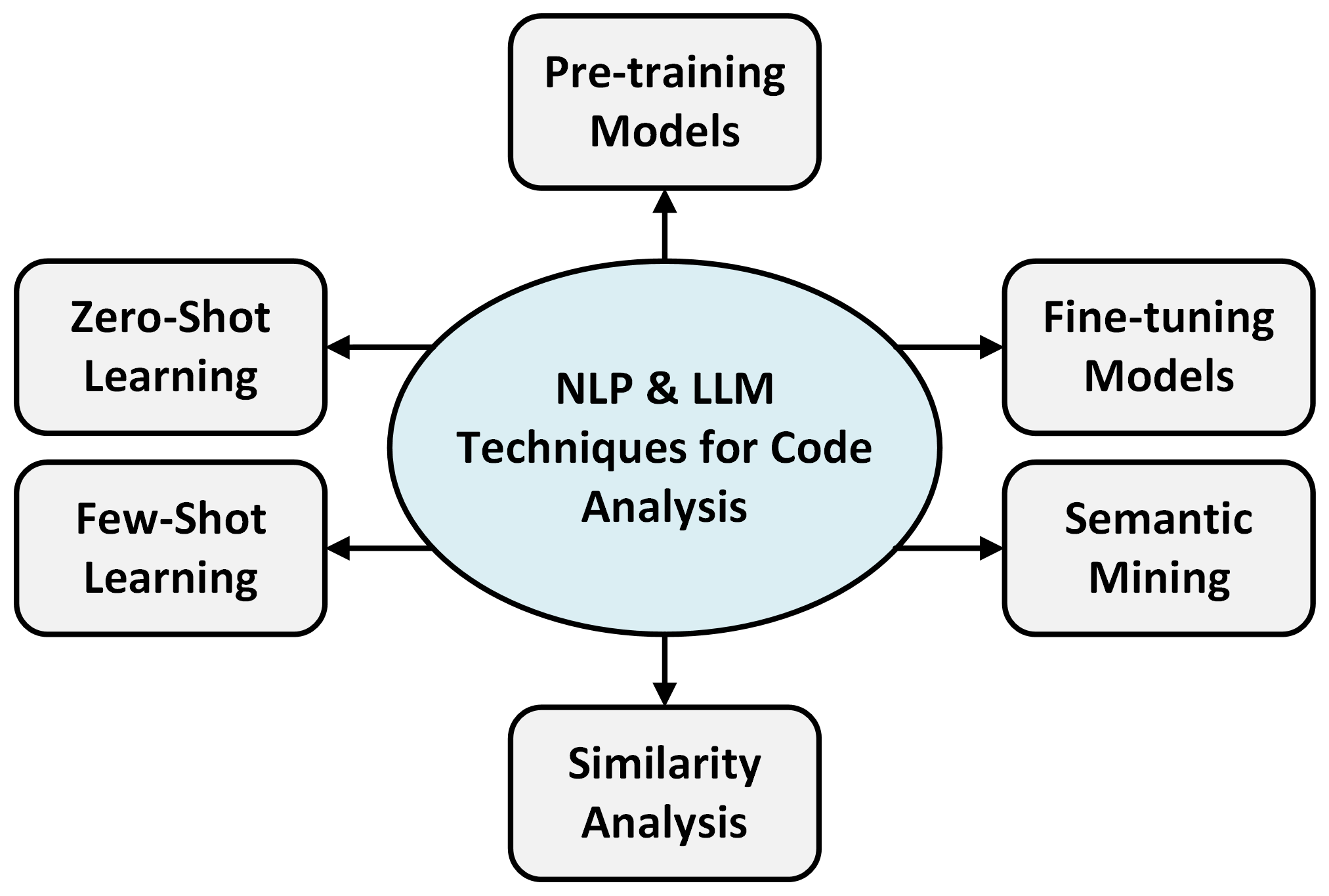}
\caption{Application of LLM models for different tasks for source code analysis}
\label{fig:x cubed graph}
\end{figure}

Another notable model is Codex \cite{149}, which powers GitHub Copilot for intelligent auto-completion, enabling developers to code more efficiently. Meanwhile, CodeBERT incorporates program structure to enhance code search and summarization tasks. These models are capable of translating code between languages, refactoring legacy code, and generating documentation with minimal human input, ultimately reducing development time and boosting productivity. Additionally, specialized tools like StarCoder \cite{150} and PolyCoder \cite{151} extend AI-assisted coding capabilities to niche programming languages, thus broadening the impact of AI technologies beyond mainstream languages. However, in Section 5, we will discuss more about the recent pre-trained LLM models. \\ 

Also, some researchers used fine-tuning techniques on LLMs after pretraining on diverse datasets like GitHub \cite{19}, CodeSearchNet \cite{31}, and Stack Overflow\cite{70}. This strategy enhances the models' ability to understand complex coding patterns and generate code for specific tasks. Additionally, they are valuable for tasks like code refactoring, performance optimization, and complex software design, offering widespread benefits across various industries.

\subsection{Why are LLMs important for code understanding?}
LLMs are essential for code understanding due to their ability to analyze, generate, and optimize code effectively. They help developers by identifying syntax errors \cite{152}, debugging issues \cite{153}, and improving code quality through suggestions for refactoring or adherence to best practices. LLMs also enhance productivity by automating repetitive tasks such as documentation generation, code completion, and creating boilerplate code. Additionally, they excel at explaining code in natural language, making complex concepts more accessible to new developers and teams.\\

Another significant advantage of LLMs is their support for advanced use cases like cross-language code translation \cite{kumar2025can, luo2024bridging, li2024few}, and testing automation \cite{sherifi2024potential, feldt2023towards}. These capabilities enable developers to work across languages more efficiently and automate testing processes, saving time and improving software quality. As LLMs advance, their role in software development is set to grow, making them essential tools for modern development workflows  \cite{127,194}. This capability bridges the gap between human and machine understanding, streamlining the development process and fostering collaboration.
\section{Applications of LLM for source code analysis}
\label{sec:applications-of-LLM-for-source-code-analysis}
LLMs demonstrate strong potential in streamlining tasks, ranging from code understanding to code summarization. By using LLMs, these models can effectively understand and generate programming languages, enhancing the efficiency of code-related tasks. Table \ref{tab:Applications of LLM for source code analysis} shows the applications of LLMs for source code analysis, based on previous works, highlighting key advancements in this field. In this Section, we discuss these LLM tasks for code analysis in more details.
\begin{table}[H]
    \centering
    \caption{Application of LLM models for different tasks for source code analysis - Code Understanding (CU), Code Disassembling (CD), Code Decompiling (CdC), Code Summarization (CS), Code Generation (CG), Comment Generation (ComG), Security Analysis (SA)}
    \resizebox{\linewidth}{!}{
    \begin{tabular}{l l}
         \hline
         App &  Explanation\\
         \hline
         CU&\begin{tabular}[c]{@{}l@{}}Analyze and comprehend the logic and functionality of code.\end{tabular}\\
         CD&\begin{tabular}[c]{@{}l@{}}Uses LLM to reverse-engineer compiled code into readable formats.\end{tabular}\\
         CdC&\begin{tabular}[c]{@{}l@{}}Converts compiled code into high-level readable formats using LLM.\end{tabular}\\
         CS&\begin{tabular}[c]{@{}l@{}}Creates concise explanations of code functionality using LLM.\end{tabular}\\
         CG&\begin{tabular}[c]{@{}l@{}}Creates executable code from natural language inputs using LLM.\end{tabular}\\
         ComG&\begin{tabular}[c]{@{}l@{}}Creates descriptive comments to explain code functionality using LLM.\end{tabular}\\
         SA&\begin{tabular}[c]{@{}l@{}}Detects vulnerabilities and ensures code security using LLM.\end{tabular}\\
         \hline
    \end{tabular}
    }
    \label{tab:Applications of LLM for source code analysis}
\end{table}
\subsection{LLM for Code Understanding}

LLMs for code understanding represent an advanced computational approach rooted in natural language processing and machine learning research \cite{131,3,4}. These models analyze the structural and semantic relationships within codebases to identify patterns, functionality, and underlying intent. Table \ref{tab:CU-CD-CdC} shows some significant works based on LLM for code understanding.  \\

In \cite{3}, authors explored the use of a LLM-based conversational UI integrated into an IDE to aid code comprehension. They found that tool benefits vary between students and professionals, with usage influenced by learning styles and AI tool experience. In \cite{4}, the survey provides an overview of the benefits of incorporating code into LLMs’ training data. It examines how code improves LLM performance and strengthens their role as intelligent agents (IAs).

They found that code training improved programming skills, reasoning, and the generation of formalized functions for diverse applications. It also enabled LLMs to interact with evaluation modules for self-improvement.\\

In \cite{5},  the authors studied the use of LLMs to support developers in code understanding. They designed an LLM-based conversational assistant that personalizes interactions based on inferred user mental states. The findings give useful insights for building LLM tools for beginners. In \cite{6}, The authors studied natural language (NL) outlines as a tool for AI-assisted software development. NL outlines, summarizes, and partitions code, enabling bidirectional updates between code and text. They found LLMs generate high-quality outlines, benefiting tasks like code understanding, maintenance, and generation. Authors introduced CodeMMLU to evaluate the code understanding capabilities of CodeLLMs using multiple-choice QA.

In \cite{7}, the authors studied the challenges in code generation using LLMs like ChatGPT and DeepSeek-Coder, focusing on the frequent misalignment of generated code with given specifications, particularly for complex programming tasks. To address this issue, they introduced a model, which combines thought-eliciting and feedback-based prompting strategies. In \cite{8}, the authors studied enhancing JavaScript source code comprehension by integrating graph alignment into LLMs. Using graph-based representations like ASTs, their model better captures structural and semantic relationships within the code.\\\\ Also other researchers \cite{9}, studied how code documentation quality affects LLMs code understanding. They found that incorrect documentation significantly hinders LLMs' comprehension, while incomplete or missing documentation has minimal impact. This suggests that LLMs rely more on code content than on documentation for understanding, highlighting the importance of accurate documentation to avoid misleading LLMs.

\begin{table*}[!htbp]
    \centering
    \caption{Models for Code Understanding (CU), Code Disassemble (CD), and Code Decompiling (CdC)}
    \resizebox{\textwidth}{!}{
    \begin{tabular}{l c c c c c c}
         \hline         Reference&Task&Challenge&Models&\makecell{Tokenizer/\\Architecture}&Dataset&\makecell{GPU\\System} \\
         \hline
         CodeMMLU \cite{10}&CU&\makecell{Evaluating the CU capabilities of\\CodeLLM using multiple-choice QA.}&\makecell{Claude-3-sonnet,\\GPT-4o,\\Meta-Llama-3-70B,\\Mixtral-8x7B,\\DeepSeek-moe-16b}&&\makecell{10,000 multiple-choice\\questions covering\\diverse programming\\domains}&\\
         \hline
         \cite{11}&CU&\makecell{ Detecting subtle inconsistencies\\between code and its NL description}&GPT-3.5 and GPT-4&&\makecell{HumanEval-X benchmark \cite{202},\\encompassing six\\programming languages:\\ \\ }&\\
         \hline
         CodeT5+\cite{12}&CU&\makecell{A flexible encoder-decoder\\LLM for code.}&&CodeT5/T5&\makecell{Multilingual code corpora,\\CodeSearchNet}&\\
         \hline
         \cite{13}&CU&\makecell{Real-world reverse engineering\\scenarios and binary code \\understanding.}&\makecell{CodeLlama,\\DeepSeek,\\CodeGen,\\Mistra,\\Vicuna,\\ChatGPT}&\makecell{CodeLlama,\\DeepSeek,\\CodeGen,\\Mistra,\\Vicuna,\\ChatGPT}&&\\
         \hline
         \cite{14}&CU&\makecell{Judging the correctness of\\provided code solutions.}&\makecell{DeepSeek,\\CodeGen}&&\makecell{APPS (Hendrycks et al., 2021)\\dataset \cite{201}}&\\
         \hline
         \cite{15}&CU&\makecell{Understanding\\long-context code.}&\makecell{Mistra,\\Vicuna,\\ChatGPT}&&RepoQA benchmark \cite{13}&\\
         \hline
         DISASLLM \cite{14}&CD&\makecell{Combining the traditional\\disassemblers with LLMs}&\makecell{-Masked next token\\prediction (MNTP)\\-Llama Model}&Llama3&&\makecell{NVIDIA\\H100 GPU,\\AMD EPYC\\ 9124 16-core}\\
         \hline
         \cite{15}&CD&\makecell{Assembly sequence\\evaluation in\\disassemble process}&\makecell{Deep Q-learning,\\GPT-3.5-turbo}&GPT&\makecell{Google Code Jam\\(GCJ) \cite{203}}&\\
         \hline
         \cite{16}&CD&\makecell{Binary Reverse\\ Engineering}&GPT-4&GPT&Google Code Jam&\\
         \hline
         Nova$^+$ \cite{18}&CD&\makecell{A generative LM\\for assembly code.}&DeepSeek-Coder&DeepSeek-Coder&\makecell{-AnghaBench \cite{204} and The Stack\\for pretraining\\-BinaryCorp-3M \cite{205} for\\fine-tuning}&\makecell{NVIDIA RTX\\A100 GPUs}\\
         \hline
         LM4Decompile \cite{20}&CdC&\makecell{Decompilation of binary\\code into high-level\\source code}&\makecell{Sequence-to-sequence\\prediction (S2S)}&DeepSeekCoder&ExeBench \cite{206}&\makecell{NVIDIA\\A100-80GB\\GPU}\\
         \hline
         DeGPT \cite{23}&CdC&\makecell{Refining decompiled C code}&GPT-3.5-turbo-0613"&GPT&\makecell{Code Contest\\dataset}&\\
         \hline
         \cite{24}&CdC&\makecell{Binary code analysis}&\makecell{Lora,\\LLM4Decompile-6.7b,\\DeepSeek-chat,\\GPT4,\\LLM4Decompile-6.7b+FAE,\\(LlamaFactory,\\FlashAttention used\\for fine-tuning)}&\makecell{Fine-tune the\\LLM4Decompile-6.7b}&&A100-SXM4-80GB\\
         \hline
         \cite{26}&CdC&\makecell{-Binary tasks\\-Binary Summarization\\-Binary function name recovery tasks}&\makecell{GPT4,\\CodeLlama-7b,\\CodeLlama-13b,\\CodeLlama-34b,\\GPT3.5,\\Claude3,\\Gemini-Pro}&GPT4&&\\
         \hline
    \end{tabular}
    }
    \label{tab:CU-CD-CdC}
\end{table*}
\subsection{LLM for Code Disassembling}
Disassembly is a particularly challenging task, especially when dealing with obfuscated executables. These executables often contain junk bytes, which are deliberately inserted to confuse disassemblers and induce errors during the analysis process \cite{16}. Table \ref{tab:CU-CD-CdC} shows some significant work based on code disassembling. In this Section, we investigate and discuss the application of LLMs for code disassembling.

In \cite{14}, the authors explored using LLMs for disassembling obfuscated executables, aiming to improve instruction boundary identification. They developed a hybrid model combining traditional disassemblers with LLMs, fine-tuned using Llama 3 8B and the LLM2Vec method. This approach highlights the potential of LLMs in enhancing traditional disassembly techniques, particularly for handling obfuscation challenges.\\

Similarly, \cite{15} introduced a model for assembly sequence evaluation in disassembly processes. This model aids in decision-making by utilizing event graphs and reinforcement learning, where LLMs play a critical role in defining the reward function. Deep Q-learning was employed with ground truth sequences, enabling improved decision-making during disassembly. This integration of LLMs demonstrates their utility in learning-based methods for evaluating and optimizing disassembly workflows. In another study, \cite{18} focused on improving the quality of decompiled C++ code by refining decompiler outputs to produce recompilable and functional source code. The authors proposed a two-step approach based on LLMs: first, iteratively fixing syntax errors to ensure recompilability and second, identifying and correcting memory errors at runtime.\\

Additionally, \cite{17} examined GPT-4's performance in binary reverse engineering through a two-phase study. The research assessed its ability to interpret human-written and decompiled code, followed by complex malware analysis. Using both automated metrics, such as BLEU scores, and manual evaluation, the study found that while GPT-4 demonstrates strong general code comprehension, its effectiveness diminishes in more intricate technical and security-focused analyses.

Moreover, \cite{18} proposed Nova, a generative language model designed specifically for assembly code. To address the challenges posed by assembly code's low information density and diverse optimizations, Nova employs a hierarchical attention mechanism for semantic understanding and contrastive learning objectives to handle optimization variations. The model demonstrated superior performance in binary code decompilation and similarity detection compared to existing methods, showcasing LLMs' growing proficiency in specialized code domains. \\

Collectively, these studies highlight the transformative potential of LLMs in code disassembly and related tasks, offering new avenues for handling obfuscation, improving decompiled outputs, and advancing binary reverse engineering.

\subsection{LLM for Code Decompiling}  

Using LLMs for code decompiling improves both comprehension and generation of code. Accurate documentation is essential for optimizing LLM performance in this process \cite{23,26}. Table \ref{tab:CU-CD-CdC} shows some significant works based on code decompiling. In this Section, we investigate and discuss the application of LLMs for code decompiling.

In \cite{20}, the authors introduced LLM4Decompile, a series of open-source models (ranging from 1.3B to 33B parameters) designed for binary code decompilation, addressing the limitations of traditional tools like Ghidra. The authors employed DeepSeekCoder, an advanced architecture specifically designed for binary code decompilation, optimizing the LLM training process to enhance code readability and executability.  Additionally, they introduced two variants: LLM4Decompile-End for direct decompilation and LLM4Decompile-Ref for refining Ghidra outputs. These advancements showcase the transformative potential of LLMs in the decompilation process, enhancing the overall efficiency and accuracy of code translation from binary to high-level languages.  \\

In \cite{23}, the authors presented DecGPT, a two-step approach using GPT-3.5-turbo-0613 to refine decompiled C code. The first step focuses on correcting syntax errors, and the second step addresses memory errors that occur during runtime. This approach demonstrates the ability of LLMs to handle common decompilation challenges and improve the quality of the decompiled code, making it more readable and functional for developers.  \\

In other work \cite{22}, authors introduced WaDec to investigate the challenges in decompiling WebAssembly (Wasm) binaries, which are compact yet difficult to analyze and interpret. This fine-tuned LLM model converts Wasm binaries into comprehensible source code. The authors found that WaDec significantly improved the decompilation of WebAssembly binaries, paving the way for more efficient analysis of this widely used format. In \cite{23}, the authors presented additional techniques for improving decompilation processes, including Self-Constructed Context Decompilation (sc2 dec), which uses in-context learning, and Fine-grained Alignment Enhancement (FAE), which aligns assembly and source code at the statement level using debugging data. Their model, fine-tuned using the LLM4Decompile-6.7B model, further enhances the decompilation process by improving the accuracy of the reconstructed source code.  

In \cite{26}, the authors proposed a framework  for binary analysis. They evaluated various LLMs, with GPT-4 as the primary model for their framework, demonstrating the potential of LLMs in binary code analysis. Alongside these advancements, tools like PYLINGUAL, a Python library for decompiling, also contribute to the decompiling landscape, offering additional resources for handling complex decompilation tasks. Basically, these studies collectively highlight the growing role of LLMs in code decompiling, showcasing their ability to improve the readability, accuracy, and efficiency of decompiled code, as well as their potential to address challenges in decompiling various binary formats. 

\subsection{LLM for Code Summarization}
Automated code summarization focuses on generating natural language summaries for code fragments, such as methods or functions \cite{28,30,33}. In general, a code summary is a brief description, typically one sentence, that plays a key role in developer documentation \cite{33}. In this Section, we investigate and discuss the application of LLM for code summarization. \\

In \cite{28}, the authors conducted a study where they distilled GPT-3.5’s code summarization abilities into smaller, more accessible models. They carefully examined the role of model and dataset sizes in the knowledge distillation process for code summarization. The authors fine-tuned a pre-trained language model to generate code summaries using genersted-examples by GPT-3.5 as training data.\

In \cite{28}, the authors studied how to simplify GPT-3.5’s code summarization abilities into smaller models. They explored the impact of model and dataset sizes in the distillation process and fine-tuned a pre-trained model to generate code summaries using samples from GPT-3.5 as training data.

\

In \cite{29}, the authors explored how LLMs interpret binary code semantics through a large-scale study. The evaluation included models like ChatGPT and Llama 2 reflecting a broad examination of LLM capabilities. Similarly, in \cite{30}, researchers investigated LLM-based code summarization, examining different prompting techniques and their impact on performance. They found simpler approaches, such as zero-shot prompting, often outperformed more complex methods, though results varied across programming languages. In particular, CodeLlama-Instruct with 7B parameters emerged as a strong performer, though challenges remained with logic programming languages.\\

In \cite{31}, the authors focused on automatic semantic augmentation of prompts. By integrating tagged semantic facts, they helped LLMs better structure their reasoning when generating code summaries. Tested on the CodeSearchNet dataset \cite{husain2019codesearchnet}, this method improved performance in both summarization and code completion tasks. Additionally, \cite{37} presented a context-aware approach, designing prompt templates for six common contexts. They created a diverse, context-focused dataset and fine-tuned two pre-trained code models to produce meaningful and inclusive code comments.\\

In \cite{38}, the authors introduced the use-seq loss function for source code summarization, prioritizing semantic similarity at the sentence level. They demonstrated its advantage over traditional categorical cross-entropy, fine-tuning models like LLaMA and evaluating them using BLEU, METEOR, and human ratings on a funcom-java-long dataset. In addition, authors \cite{42}, proposed improving code-summary alignment using a multitask learning approach. They developed three summary-specific tasks:  masked language modeling (MLM) \cite{bitton2021data} , and unidirectional language modeling (ULM), action word prediction (AWP), enhancing the encoder's effectiveness in creating inclusive and meaningful summaries.\\

\subsection{ LLM for Code Generation}
LLMs allow automatic code generation from natural language inputs \cite{55,57}.  Table \ref{tab:Recent impressive models for code generation} shows some significant works based for code generation. In this section, we investigate and discuss the application of LLM for code generation.\\

In  \cite{50}, the authors studied the non-determinism in code generation, specifically examining the variability in its outputs when identical prompts were used. This research aimed to quantify this non-determinism and understand its impact on the reliability and reproducibility of LLMs in software engineering. By comparing the semantic, syntactic, and structural differences in the generated code under varying temperature settings, they found significant variability. This study highlighted the challenges that non-determinism poses for developers and researchers when using ChatGPT in coding tasks.\\

In  \cite{53}, the authors analyzed social biases in code generated by LLMs like GPT-4 and GPT-3.5-turbo, focusing on age, gender, and race. They found 13.47\% to 49.10\% of outputs displayed gender bias using a novel bias testing framework. To address this, they evaluated mitigation strategies such as zero-shot, one-shot, few-shot, and Chain-of-Thought prompts \cite{li2025structured}, both with and without feedback refinement. While direct prompt engineering had limited success, feedback-driven strategies significantly reduced bias, cutting GPT-4's bias rate from 59.88\% to 4.79\%. This highlights the importance of execution feedback in reducing biases in code generation.\\

In  \cite{55}, the authors created FlowGen, a framework that uses LLM agents to simulate software process models. These agents refine code through reasoning and prompts. Tested with GPT-3.5 on benchmarks like HumanEval, adding CodeT to FlowGenScrum achieved the best Pass@1 scores.\\

In  \cite{63}, the research also focused on TDD for code generation, they investigated if and how Test-Driven Development(TDD) can be incorporated into AI-assisted code-generation processes. For experiment they used GPT-4, Llama 3 methods based on MBPP and HumanEval datasets. Further exploration of TDD in LLM-based code generation suggests that combining traditional software engineering practices with AI-driven approaches can significantly improve the quality of generated code. Researchers have pointed out that incorporating a testing-first approach, such as TDD, helps in ensuring that the code not only meets functional requirements but also adheres to desired performance and stability metrics.

\subsubsection{Code generation and security}
There are some researches that focused on the security of code generation by LLMs. For instance, in \cite{197} focused on evaluating the security of AI-generated code by introducing CodeSecEval, a dataset with 180 samples covering 44 critical vulnerability types. They assessed LLMs on both code generation and repair tasks, revealing that these models often produce insecure code due to training on unsanitized open-source data. \\

Also, in \cite{198}, the authors proposed a framework to assess LLM-generated code for security and functionality. They introduced CWEval-Bench, a multilingual benchmark with 119 tasks covering 31 CWE types. Their evaluation showed that many functionally correct code outputs are insecure.\\

\subsubsection{Code generation and explanation}
LLMs not only are revolutionizing code generation, but also are quite useful in generating clear explanation for code. Tools based on LLM technology, enable developers to receive code summary and explanation only by highlighting the specific part of the code without any need to query a prompt. Studies have indicated that such tools improve task completion rates, especially for experienced developers, by reducing the load related to manual searches through documentation  \cite{1}.

\subsection{LLM for Comment Generation}
Code review is a key part of modern software development, helping to improve both the quality of systems and the skills of developers. Table \ref{tab:Recent impressive models for comment generation} shows some significant works based for comment generation.  Recently, studies have explored LLM-based methods to automatically generate review comments, making the process more efficient. In this section, we investigate and discuss the application of LLMs for comment generation \cite{69,72,73,74,75,76,77}\\

In \cite{69}, the authors studied the integration of LLMs into the code review process, focusing on their ability to generate review comments and their acceptance by human reviewers. They conducted a large-scale empirical user study within two organizations—Mozilla (open-source) and Ubisoft (closed-source)—to evaluate the effectiveness and reception of LLM-generated comments in real-world settings. The model used for this study was based on OpenAI’s GPT-4. Accepted comments generated by LLMs were just as likely to prompt future revisions of the modified patch as human-written comments, indicating that LLM-generated feedback can play a valuable role in the code review process. This highlights the potential for LLMs to automate and enhance the efficiency of code reviews.  \\

In \cite{71}, the authors fine-tuned Llama models with QLoRA to improve code review comments and explored enhancing inputs with function call graphs and summaries, highlighting the value of added code context. This work emphasizes the potential of augmenting LLMs with additional code context to improve comment quality.  \\

In \cite{73}, the authors proposed SCCLLM, which retrieves top-k code snippets for in-context learning to generate smart contract comments. Their study highlights LLMs' potential in automating code review and comment generation.

\subsection{LLM for Security Code Analysis}
LLMs generate insecure code by default, up to 65\% of the generated codes contain vulnerabilities. Table \ref{tab:frameworks-comparisom-m} shows some significant works for LLM in security code. However, a skilled engineer can manually guide an LLM to generate a 100\% secure code. To gain the best possible results from an LLM prompts need to be optimized. The prompts need to be both interpretable for an AI agent to comprehend and precise to gain better results. Moreover, breaking a complex prompt into multiple simple prompts can generally lead to higher performance  \cite{80}.\\

Code (Security) analysis is the process of detecting and fixing vulnerabilities in a code to prevent attacker's from breaching the system. Code analysis could be either static or dynamic. In static code analysis, logic and syntax faults are handled. On the other hand, in dynamic code analysis, the code is analyzed during the run time to solve memory problems, bottlenecks and so on  \cite{81}.

LLMs can enhance security code analysis by detecting vulnerabilities early in the development phase \cite{82}. They help identify bugs and insecure coding practices, such as weak authentication or improper data handling, and provide recommendations for improving code security \cite{156}.

For example, some researchers In \cite{cheng2024security} focused on evaluating the security vulnerabilities of Large Language Model-based Code Completion Tools (LCCTs), specifically GitHub Copilot and Amazon Q. They investigated two primary threats: jailbreaking attacks, which bypass built-in safety mechanisms to generate harmful or unethical code, and training data extraction attacks, where sensitive information from the model's training data is unintentionally exposed. The study serves as a call for greater scrutiny in deploying LLM-powered code completion tools, ensuring they adhere to stricter security and ethical guidelines.\\

Also in \cite{kharma2025security}, The authors focused on evaluating the security and quality of code generated by Large Language Models (LLMs) across multiple programming languages. They analyzed models like GPT-4o, Claude-3.5, Gemini-1.5, Codestral, and LLaMA-3 using a dataset of 200 programming tasks, categorized into six groups. Their primary goal was to assess how well these models adhere to modern security practices and whether they produce maintainable, high-quality code. Their findings indicate that while LLMs are effective in automating code generation, their security performance varies significantly across languages.

\subsubsection{Frameworks}
~~~~In order to securely generate a source code using an LLM, some frameworks have been proposed, including SecCode  \cite{82}, SALLM  \cite{79} and LLMSecGuard  \cite{78} to do the prompt engineering task in an automatic manner.

SecCode operates in 3 main stages, code generation, code vulnerability detection (code analysis) and fixing, and code security refinement. The process begins when the user enters prompt $p_1$ and code $C$ is generated using an LLM model like ChatGPT. The generated code is not necessarily secure. In the second stage, an automated prompt $P_2$ is generated by the system leading to the detection and fixing vulnerabilities using a rewarding system called Encouragement Prompting (EP). EP rewards the system if a vulnerability is found and fixed. On the other hand, EP penalizes the system if some new vulnerabilities are introduced or the existing vulnerabilities remain after fixing process \cite{82}. 

Moreover, after initial fixes, the code undergoes further security checks using CodeQL. If CodeQL finds additional hidden vulnerabilities, a prompt $p3$ is automatically generated leading to further code security enhancement by LLM and this process repeats over and over until CodeQL can't find anymore vulnerabilities  \cite{82}. 

SALLM includes 3 components, a dataset of python prompts, a code generation and repair module, and a security evaluation system on model's performance. At first, a dataset of security-centric is created by collecting prompts from different sources. These prompts are used to generate code by feeding them into the LLM, which returns the top $k$ results for each prompt. After that, SALLM evaluates the security of the generated code using static and dynamic analysis. Finally, SALLM iteratively refines the code until no vulnerability is found  \cite{79}.

LLMSecGuard takes the user's prompt, enhances it, and forwards it to an LLM to generate the code. Then, it sends the code static code analysis engines including Semgrep and Weggli to measure how secure the code is. This process is repeated throughout the iterations until certain conditions are met \cite{78}.

\subsubsection{Security and code Documentation}
The quality of code documentation is another crucial factor influencing the effectiveness of LLMs assisting developers. Accurate and detailed documentation improves LLM performance, while incorrect or misleading documentation can significantly hinder code understanding task. Surprisingly, the absence of documentation often has less of a negative impact compared to flawed comments or descriptions, meaning that the quality of the documentations and comments should be strictly preserved  \cite{7}.
 
We can observe that these frameworks work along with existing LLMs and code analysis systems. So, they do not inherently contain these modules. In table \ref{tab:frameworks-comparisom-m}, a comparison between some recent security analysis frameworks is shown:
\begin{table}[h]
    \centering
    \caption{Recent Security Code Generation frameworks}
    \resizebox{0.5\textwidth}{!}{
    \begin{tabular}{l c c c c c c}
        \hline
        Framework & Models & \makecell{Code\\Analysis\\Tool} & Dataset & \makecell{Language(s)}\\
        \hline
        SecCode \cite{82} &\makecell{GPT-3.5\\DeepSeek-Coder}&CodeQL&\makecell{LLMSecEval \cite{211}\\SecurityEval \cite{210}\\Holmes \cite{86}}&\makecell{C\\Python}\\
        \hline

        SALLM \cite{79} &\makecell{CodeGen models\\StarCoder\\GPT models}&CodeQL&\makecell{LLMSecEval \cite{211}\\SecurityEval \cite{210}\\SALLM \cite{79}}&\makecell{C\\Python}\\
        \hline

        LLMSecGuard \cite{78} &Llama2&\makecell{Semgrep\\Weggli}&CyberSecEval \cite{212}&Java\\
        \hline
    \end{tabular}
    }
    \label{tab:frameworks-comparisom-m}
\end{table}





\begin{table*}[!htbp]
\caption{Recent models for code summarization}
\resizebox{\textwidth}{!}{ 

\begin{tabular}{l c c c c c c}
\hline
Ref                                                                                        & Year                     & Challenge                                                                                        & Models                                                                                                         & \begin{tabular}[c]{@{}l@{}}Tokenizer/\\architecture\end{tabular}                                                                & Dataset                                                                                                                                          & GPU System                                                                                                              \\ \hline
\multirow{3}{*}{\cite{28}}                                                          & 2024                    & \begin{tabular}[c]{@{}l@{}}-Fine-tuning\\for code \\summarization\end{tabular}                                                               &                                                                                                                & GPT                                                                                    &                                                                                                                                                  &                                                                                                                         \\  
                                                                                                 &  &                                                                                          &                                                                                              &                                                                   &                                                                                                                                 &                                                                                                        \\ 
                                                                                                 &                          &                                                                &    GPT-3.5                                                                                                            &                                                                                       &        \begin{tabular}[c]{@{}l@{}}-Collected\\ summaries\\ for 2.15m Java\end{tabular}                                                                                                                                          &                                                                                                                         \\ \hline
\multirow{5}{*}{ \cite{29}}                                                          & \multirow{5}{*}{2024}    & \begin{tabular}[c]{@{}l@{}}LLMs' understanding \\of binary code\end{tabular}                                             & - GPT-4,                                                                                                       &                                                                                        & \begin{tabular}[c]{@{}l@{}}44 open-access software project \\ \end{tabular}                                                               & \begin{tabular}[c]{@{}l@{}}NVIDIA\\A100 GPU\end{tabular}                                                                                                         \\ 
                                                                                                 &                          &                                                                                                  & - ChatGPT (GPT-3.5),                                                                                           &                                                                                        &                                                                                                                                                  &                                                                                                                         \\  
                                                                                                 &                          &                                                                                                  & - Llama 2                                                                                                      &                                                                                        &                                                                                                                                                  &                                                                                                                         \\  
                                                                                                 &                          &                                                                                                  & - Code Llama                                                                                                   &                                                                                        &                                                                                                                                                  &                                                                                                                         \\  
                                                                                                 &                          &                                                                                                  & - BinT5                                                                                                        &                                                                                        &                                                                                                                                                  &                                                                                                                         \\ \hline
\multirow{5}{*}{  \cite{29}}                                                          & \multirow{5}{*}{2024}    & \begin{tabular}[c]{@{}l@{}}LLM evaluation\\for code summary\end{tabular}                                                  & \begin{tabular}[c]{@{}l@{}}CodeBERT \\ -GraphCodeBERT\end{tabular}                                             & -                                                                                      & CodeSearchNet                                                                                                                                    &                                                                                                                         \\ 
                                                                                                 &                          &                                                                                                  & -Polyglot CodeBERT                                                                                             &                                                                                        &                                                                                                                                                  &                                                                                                                         \\  
                                                                                                 &                          &                                                                                                  & \begin{tabular}[c]{@{}l@{}}-Polyglot \\ GraphcodeBERT\end{tabular}                                             &                                                                                        &                                                                                                                                                  &                                                                                                                         \\  
                                                                                                 &                          &                                                                                                  & - CodeT5                                                                                                       &                                                                                        &                                                                                                                                                  &                                                                                                                         \\  
                                                                                                 &                          &                                                                                                  & -CodeT5+                                                                                                       &                                                                                        &                                                                                                                                                  &                                                                                                                         \\ \hline
\multirow{4}{*}{ \cite{30}}                                                          & \multirow{4}{*}{2024}    & \begin{tabular}[c]{@{}l@{}}A heuristic\\for ChatGPT's \\prompts.\end{tabular}                                                & NCS,                                                                                                           &                                                                                        & CSN-Python dataset                                                                                                                               &                                                                                                                         \\  
                                                                                                 &                          &                                                                                                  & CodeBERT,                                                                                                      &                                                                                        & CodeSearchNet                                                                                                                                    &                                                                                                                         \\  
                                                                                                 &                          &                                                                                                  & and CodeT5                                                                                                     &                                                                                        &                                                                                                                                                  &                                                                                                                         \\  
                                                                                                 &                          &                                                                                                  & ChatGPT                                                                                                        &                                                                                        &                                                                                                                                                  &                                                                                                                         \\ \hline
\cite{33}                                                                           & 2024                     & \begin{tabular}[c]{@{}l@{}}Exploring\\ChatGPT's \\summarization  \end{tabular}                                                              & ChatGPT                                                                                                        &                                                                                        &                                                                                                                                                  &                                                                                                                         \\ \hline

\multirow{2}{*}{\cite{36}}                                                          & \multirow{2}{*}{2024}    & \begin{tabular}[c]{@{}l@{}}Fine-tuning for\\code summarization\end{tabular}                                              & \multirow{2}{*}{Llama 7B}                                                                                      & \multirow{2}{*}{Llama 7B}                                                              & \multirow{2}{*}{Funcom-java-long dataset.}                                                                                                       & {\begin{tabular}[c]{@{}l@{}}AMD 5900X CPU, \\ 128GB memory, \\and two Nvidia \\ A5000 GPU\end{tabular}} \\
                                                                                                 &                          &                                                                                                  &                                                                                                                &                                                                                        &                                                                                                                                                  &                                                                                                                         \\ \hline
\cite{37}                                                                           & 2024                     & \begin{tabular}[c]{@{}l@{}}Fine-tuning for\\code summarization\end{tabular}                                                               & Fine-tune the Llama2-7B                                                                                        & Llama2-7B m                                                                            & Python code dataset                                                                                                                               & \begin{tabular}[c]{@{}l@{}}NVIDIA\\A100-SXM \\ 40GB-RAM \\GPU\end{tabular}                                                 \\ \hline
\multirow{3}{*}{\begin{tabular}[c]{@{}l@{}}\cite{40}\end{tabular}} & \multirow{3}{*}{2024}    & \begin{tabular}[c]{@{}l@{}}Fine-tuning for\\code summarization\end{tabular}                                              & \multirow{3}{*}{CodeLlama-7B-instruct}                                                                         & \multirow{3}{*}{\begin{tabular}[c]{@{}l@{}}CodeLlama-7B-instruct \\ mode\end{tabular}} & \begin{tabular}[c]{@{}l@{}}http://2https//www.kaggle.com/datasets\\ /pelmers/github-repository-metadata\\ -with-5-stars/versions/83\end{tabular} & \begin{tabular}[c]{@{}l@{}}NVIDIA V100\\16Gb GPU\end{tabular}                                                                                  \\ 
                                                                                                 &                          &                                                                                                  &                                                                                                                &                                                                                        & \multirow{2}{*}{https://clang.llvm.org/}                                                                                                         &                                                                                                                         \\
                                                                                                 &                          &                                                                                                  &                                                                                                                &                                                                                        &                                                                                                                                                  &                                                                                                                         \\ \hline
\multirow{3}{*}{ \cite{42}}                                                          & \multirow{3}{*}{2024}    & \begin{tabular}[c]{@{}l@{}}-Automatic code \\summarization\end{tabular}                                                    & Llama-70B                                                                                                      & \multirow{3}{*}{}                                                                      & \multirow{2}{*}{\begin{tabular}[c]{@{}l@{}}Ericsson and open-source projects \\ (Guava and Elasticsearc)\end{tabular}}                           & \multirow{2}{*}{Tesla A100 GPU}                                                                                         \\ 
                                                                                                 &                          &                                                                                                  & \multirow{2}{*}{\begin{tabular}[c]{@{}l@{}}CodeLlama \\ Mixtral LLM\end{tabular}}                              &                                                                                        &                                                                                                                                                  &                                                                                                                         \\  
                                                                                                 &                          & \begin{tabular}[c]{@{}l@{}}-Conducted\\ experiments on \\ an Ericsson\\ software project\end{tabular} &                                                                                                                &                                                                                        &                                                                                                                                                  &                                                                                                                         \\ \hline
\begin{tabular}[c]{@{}l@{}}\cite{46}\end{tabular}                    & 2024                     & \begin{tabular}[c]{@{}l@{}}Evaluate\\code summary\\similarity\end{tabular}                                                                   & \begin{tabular}[c]{@{}l@{}}roBERTa, \\ MPNet\end{tabular}                                                      &                                                                                        & \begin{tabular}[c]{@{}l@{}}2 million\\code summaries\\(pre-0t).\end{tabular}                                                                                                                & \begin{tabular}[c]{@{}l@{}}d 4 NVIDIA \\A100 GPUs, \\ each with \\80 GB \\of VRAM\end{tabular}                                \\ \hline
\end{tabular}
\label{tab:Recent impressive models for code summarization}
}
\end{table*}

\clearpage


\begin{table*}[!htbp]
\clearpage
\caption{Recent models for Code Generation}

\resizebox{\textwidth}{!}{ 
\begin{tabular}{l l l l l l l}
\hline
Reference                                & Year                  & Challenge                                                                                                   & Models                                                                                                                                                                                 & \begin{tabular}[c]{@{}l@{}}Tokenizer\\/architecture\end{tabular}                                         & Dataset                                                                                                                                                                 & GPU System                                                                                         \\ \hline
\cite{50}                   & 2024                  & \begin{tabular}[c]{@{}l@{}}Code generation\end{tabular}                                                                                        & ChatGPT                                                                                                                                                                                & GPT/ChatGPT                                                     & \begin{tabular}[c]{@{}l@{}}CodeContests \cite{64}, \\ APPS, and \\HumanEval\end{tabular}                                                                 &                                                                                                    \\ \hline
\multirow{2}{*}{\cite{51}}  & \multirow{2}{*}{2024} & \begin{tabular}[c]{@{}l@{}}Prompt engineering\\in code bias\end{tabular}                                                                 & \multirow{2}{*}{\begin{tabular}[c]{@{}l@{}}PALM-2/GPT\end{tabular}}                                    &                                                                 &                                                                                                                                                                         &                                                                                                    \\  
                                         &                       &                                                                                                                  &                                                                                                                                                                                        &                                                                 &                                                                                                                                                                         & \multirow{4}{*}{}                                                                                  \\ \hline
\multirow{3}{*}{\cite{50}}  & \multirow{3}{*}{2023} & \multirow{3}{*}{\begin{tabular}[c]{@{}l@{}}Hallucinations in \\ code generation\end{tabular}}         & ChatGPT                                                                                                                                                                                & \multirow{3}{*}{-}                                              & HumanEval                                                                                                                                                               &                                                                                                    \\ 
                                         &                       &                                                                                                                  & CodeLLama-7B5,                                                                                                                                                                         &                                                                 & \multirow{2}{*}{HALLUCODE \cite{52}}                                                                                                                   &                                                                                                    \\ 
                                         &                       &                                                                                                                  & DeepSeek-Coder-7B6                                                                                                                                                                     &                                                                 &                                                                                                                                                                         &                                                                                                    \\ \hline
FlowGen \cite{53}       & 2024                  & \begin{tabular}[c]{@{}l@{}}Emulating software \\ with LLMs.\end{tabular}                            & GPT-3                                                                                                                                                                                  & GPT                                                             & \begin{tabular}[c]{@{}l@{}}HumanEval,\\HumanEval-ET, \\ MBPP (Mostly\\ Basic Python\\Programming),\end{tabular}                                                          &                                                                                                    \\ \hline
TiCoder \cite{54}       & 2024                  & \begin{tabular}[c]{@{}l@{}}Improve code\\suggestion \\accuracy.\end{tabular}                                                                                & \begin{tabular}[c]{@{}l@{}}OpenAI code-davinci-002, \\ text-davinci-003, \\ OpenAI GPT-3.5-turbo, \\ GPT-4-turbo, \\ GPT-4-32k, \\ Salesforce CodeGen-6B,\\ CodeGen2.5-7B\end{tabular} & -                                                               & \begin{tabular}[c]{@{}l@{}}Two Python datasets,\\ HumanEval \\ and MBPP\end{tabular}                                                                                      &                                                                                                    \\ \hline
ClarifyGPT \cite{55}    & 2024                  & \begin{tabular}[c]{@{}l@{}}Enhance code\\generation by \\identifying\\ambiguities\end{tabular}                     & GPT-4 and ChatGPT                                                                                                                                                                      & GPT                                                             & \begin{tabular}[c]{@{}l@{}}MBPP-sanitized , \\ HumanEval, \\ HumanEval-ET \\and MBPP-ET,\\ CoderEval \cite{208}\end{tabular}                  & -                                                                                                  \\ \hline
\multirow{2}{*}{\cite{56}}  & \multirow{2}{*}{2024} & \begin{tabular}[c]{@{}l@{}}Code generation with \\ image recognition.\end{tabular}                           & \multirow{2}{*}{GPT-4}                                                                                                                                                                 & \multirow{2}{*}{GPT}                                            & \multirow{2}{*}{}                                                                                                                                                       & \multirow{2}{*}{}                                                                                  \\
                                         &                       &                                                                                                                  &                                                                                                                                                                                        &                                                                 &                                                                                                                                                                         &                                                                                                    \\ \hline
\cite{57}                   & 2024                  & \begin{tabular}[c]{@{}l@{}}LLM search for\\code generation.\end{tabular}                                                                                  & OpenAI’s GPT-4o-mini a.                                                                                                                                                                & \begin{tabular}[c]{@{}l@{}}OpenAI’s \\ GPT-4o-mini\end{tabular} & \begin{tabular}[c]{@{}l@{}}MBPP+,\\HumanEval+\\  28  ,\\ and \\LiveCodeBench \\  MBPP   \end{tabular} &                                                                                                    \\ \hline
\begin{tabular}[c]{@{}l@{}}Deceptprompt\\ \cite{59}\end{tabular}  & 2023                  & \begin{tabular}[c]{@{}l@{}}Instructions\\for code\end{tabular}                                          & \begin{tabular}[c]{@{}l@{}}1) CodeLlama7B \\ 2) StarChat-15B \\ (fine-tuned on\\StarCoder15B ); \\ 3) WizardCoder-3B and\\ 4) WizardCoder15B. T\end{tabular}                           &                                                                 & \begin{tabular}[c]{@{}l@{}}A dataset that\\covers 25 different \\CWE types\end{tabular}                                                                                 &                                                                                                    \\ \hline
\begin{tabular}[c]{@{}l@{}}NL2ProcessOps\\ \cite{66}\end{tabular} & 2024                  & \begin{tabular}[c]{@{}l@{}}Retrieval Augmented \\ Generation (RAG) to \\ streamline code\\generation.\end{tabular} & \begin{tabular}[c]{@{}l@{}}GPT-4\\(gpt-4-0125-preview)\end{tabular}                                                                                                                                                             & GPT-4                                                           & GitHub Copilot                                                                                                                                                          &                                                                                                    \\ \hline
\cite{67}                   & 2024                  & \begin{tabular}[c]{@{}l@{}}Improved\\cost-accuracy\\trade-offs.\end{tabular}                             & \begin{tabular}[c]{@{}l@{}}Codegen-mono, \\ WizardCoder-V1.0, \\ WizardCoder-Python-V1.0\end{tabular}                                                                                  &                                                                 & \begin{tabular}[c]{@{}l@{}}HumanEvalm, \\ MBPP-sanitized, \\ APPS-test\end{tabular}                                                                                     & \begin{tabular}[c]{@{}l@{}}NVIDIA\\Geforce RTX \\ 3090 GPUs, \\ each with\\24GB of VRAM\end{tabular} \\ \hline
\cite{68}                   &                       & \begin{tabular}[c]{@{}l@{}}Code generation for \\ audio programming.\end{tabular}                                & \begin{tabular}[c]{@{}l@{}}MaxMSP,\\gpt-4-0125-preview\end{tabular}                                                                                                                                                             &                                                                 &                                                                                                                                                                         &                                                                                                    \\ \hline
\end{tabular}
\label{tab:Recent impressive models for code generation}
}
\end{table*}


\clearpage
\begin{table*}[!htbp]
\centering
\caption{Models for pre-training and LLMs}

\resizebox{0.97\textwidth}{!}{ 
\begin{tabular}{l c c c c}
\hline
Reference                                                       & Details                                                                                                                                                & Architecture/Pre-trained Model & Pre-trained Domain/Dataset                                                                                                                     & GPU System                                                                                               \\ \hline
\multirow{2}{*}{AnchorCoderP \cite{19}} & \begin{tabular}[c]{@{}l@{}}A pre-trained model\\based on Llama \end{tabular}                                                                                                   & \multirow{2}{*}{CodeLlama-7B}  & \multirow{2}{*}{CodeSearchNet}                                                                                                                                    & \begin{tabular}[c]{@{}l@{}}4 NVIDIA A40\\48GB GPUs \end{tabular}                                                                    \\
                                                          &                                                                                                                                                        &                                &                                                                                                                                                                   &                                                                                                            \\ \hline
  MONOCODER \cite{159}                  & \begin{tabular}[c]{@{}l@{}}Smaller LMs for\\specific domains\end{tabular}                                               & GPT-3.5 r                      & \begin{tabular}[c]{@{}l@{}}HPC-specific dataset \\ \end{tabular}                                                                       & \begin{tabular}[c]{@{}l@{}}4 NVIDIA\\A40 48GB GPUs \end{tabular}                                                                                     \\ \hline
  CommitBART \cite{160}                  & \begin{tabular}[c]{@{}l@{}}Based on GitHub\\commits \end{tabular} & BART {[}27{]} architecture     & \begin{tabular}[c]{@{}l@{}}GitHub commits/ \\  7 programming languages \\ \end{tabular}   &                                                                                                            \\ \hline
 \cite{161}                               & \begin{tabular}[c]{@{}l@{}}Improving logic and \\instruction-following.\end{tabular}                         & GPT2, LLaMA (Tiny-Llama v1.1)  & \begin{tabular}[c]{@{}l@{}}Ten programming languages \\ \\  and other datasets\\  (Wikipedia) \end{tabular} & 8 H100 GPUs                                                                                                \\ \hline
  \cite{162}                              & \begin{tabular}[c]{@{}l@{}}Code-aware\\program repair\end{tabular}                                                                                                             & GPT                            & \begin{tabular}[c]{@{}l@{}}A very large software codebase\\ 4.04 million methods \\ from 1,700 open-source projects\end{tabular}                                  & \begin{tabular}[c]{@{}l@{}}one 56-core server with one \\ NVIDIA TITAN V and\\  three Xp GPUs\end{tabular} \\ \hline
TreeBERT \cite{163}                  & \begin{tabular}[c]{@{}l@{}}An LLM for\\programming\\language\end{tabular}                                                                                                       & TreeBert/Bert                  &   Python and Java corpus (CuBERT)                                                                                                                      &                                                                                                            \\ \hline
NatGen \cite{164}                    & \begin{tabular}[c]{@{}l@{}}An LLM for\\code generation\end{tabular}                                                                                                             & CodeT5                         & CodeSearchN                                                                                                                                                       & \begin{tabular}[c]{@{}l@{}}2 Nvidia\\GeForce RTX\\3090 GPUs  \end{tabular}                                                                            \\ \hline
\cite{165}                            & \begin{tabular}[c]{@{}l@{}}Evaluating \\pre-trained \\transformers\end{tabular}                                         & seBERT                         & \begin{tabular}[c]{@{}l@{}}Stack Overflow, Jira, GitHub,\\ \end{tabular}                                                               & \begin{tabular}[c]{@{}l@{}}8x NVIDIA\\Tesla V100\\32G GPUs \end{tabular}                                                                             \\ \hline
\end{tabular}
\label{tab:Some impressive models based on pre-training and LLMs}
}
\end{table*}

\begin{table*}[!htbp]
\centering
\caption{Recent Models for comment generation}

\resizebox{0.97\textwidth}{!}{ 
\begin{tabular}{l l l l l l l}
\hline
Reference                                                                                          & Year                  & Challenge/Goal                                                                                                                                           & Models                                                                                                                                                & \begin{tabular}[c]{@{}l@{}}Tokenizer\\/architecture \end{tabular}& Dataset                                                                                                                  & GPU System                                                                                              \\ \hline
\begin{tabular}[c]{@{}l@{}}\\RevMate \cite{69}\\ \end{tabular}                      & 2024                  & \begin{tabular}[c]{@{}l@{}}Comment\\ generation\end{tabular}                                                                         & GPT4o                                                                                                                                                 & GPT                     &                                                                                                                          &                                                                                                         \\ \hline
\begin{tabular}[c]{@{}l@{}}AUTOGENICS \cite{70} \\ \end{tabular} & {2024} & {\begin{tabular}[c]{@{}l@{}}Inline\\comment\\generation\end{tabular}} & Gemini 1.5 Pro & & \begin{tabular}[c]{@{}l@{}}400 code snippets \\ (200 Python + 200 Java)\end{tabular}                    &                                                                                        \\ 
                                                                                                   &                       &                                                                                                                                                          & \multirow{2}{*}{- GPT-4}                                                                                                                              &                         &                                                                                                                          &                                                                                                         \\
                                                                                                   &                       &                                                                                                                                                          &                                                                                                                                                       &                         &                                                                                                                          &                                                                                                         \\ \hline
\multirow{6}{*}{\cite{73}}                                                            & \multirow{6}{*}{2024} & {\begin{tabular}[c]{@{}l@{}}Improving\\review \\comments.\end{tabular}}    & \begin{tabular}[c]{@{}l@{}}Llama  models \end{tabular} & \multirow{6}{*}{GPT}    & \multirow{6}{*}{CodeReviewer dataset \cite{209}}                                                                                    & \multirow{6}{*}{\begin{tabular}[c]{@{}l@{}}NVIDIA RTX\\5000 GPU\\machine with\\16 GB VRAM\end{tabular}} \\ 
                                                                                                   &                       &                                                                                                                                                          & \multirow{2}{*}{-GPT-3.5}                                                                                                                             &                         &                                                                                                                          &                                                                                                         \\
                                                                                                   &                       &                                                                                                                                                          &                                                                                                                                                       &                         &                                                                                                                          &                                                                                                         \\ 
                                                                                                   &                       &                                                                                                                                                          & \multirow{2}{*}{QLoRA technique to fine-tune}                                                                                                         &                         &                                                                                                                          &                                                                                                         \\
                                                                                                   &                       &                                                                                                                                                          &                                                                                                                                                       &                         &                                                                                                                          &                                                                                                         \\ 
                                                                                                   &                       &                                                                                                                                                          & - CodeT5                                                                                                                                              &                         &                                                                                                                          &                                                                                                         \\ \hline
\multirow{2}{*}{\cite{74}}                                                            & \multirow{2}{*}{2024} & {\begin{tabular}[c]{@{}l@{}}Multi-intent\\comment \\generation\end{tabular}}                                   & \multirow{2}{*}{Codex model code-davinci-002.}                                                                                                        & \multirow{2}{*}{}       & \multirow{2}{*}{\begin{tabular}[c]{@{}l@{}}Funcom and TLC  \end{tabular}} & \multirow{2}{*}{}                                                                                       \\
                                                                                                   &                       &                                                                                                                                                          &                                                                                                                                                       &                         &                                                                                                                          &                                                                                                         \\ \hline
\multirow{2}{*}{\begin{tabular}[c]{@{}l@{}}SCCLLM \cite{73} \\ \end{tabular}}     & \multirow{2}{*}{2024} & {\begin{tabular}[c]{@{}l@{}}Automatic\\comment \\generation\end{tabular}}                           & CodeBERT                                                                                                                                              & \multirow{2}{*}{}       & \multirow{2}{*}{\begin{tabular}[c]{@{}l@{}}A large corpus from Etherscan.io\end{tabular}}                             & \multirow{2}{*}{\begin{tabular}[c]{@{}l@{}}GeForce\\RTX4090 GPU\\ (24GB\\graphic memory)\end{tabular}}    \\ 
                                                                                                   &                       &                                                                                                                                                          & \begin{tabular}[c]{@{}l@{}}    \\       \end{tabular}           &                         &                                                                                                                          &                                                                                                         \\ \hline
\end{tabular}
\label{tab:Recent impressive models for comment generation}
}
\end{table*}
\clearpage

\section{Pre-training for domain adaptation on code analysis}
\label{sec:Pre-training for Domain Adaptation on code
analysis}

Pre-training involves initially training a model on a large, general-purpose dataset, such as predicting the next word in a sequence. Following pre-training, the model undergoes fine-tuning on a smaller, domain-specific dataset to optimize its performance on specialized tasks like text generation. Figure 4 illustrates the pre-training pipeline for LLMs. As indicated in Table \ref{tab:Some impressive models based on pre-training and LLMs} and corroborated by prior research, datasets like CodeSearchNet, HPCorpus, and commit data significantly boost the model’s ability to understand and perform in specific domains, such as code generation. For instance, in \cite{zhang2024anchor}, the authors introduced AnchorCoder, a novel approach designed to enhance LLMs for code generation while minimizing computational resource requirements. AnchorCoderP \cite{zhang2024anchor}, pre-trained on CodeSearchNet from scratch, was evaluated across various language modeling tasks, revealing attention weight sparsity patterns where information consolidated at distinct anchor points.

\begin{figure}[h!]
  \centering
  \includegraphics[width=\linewidth]{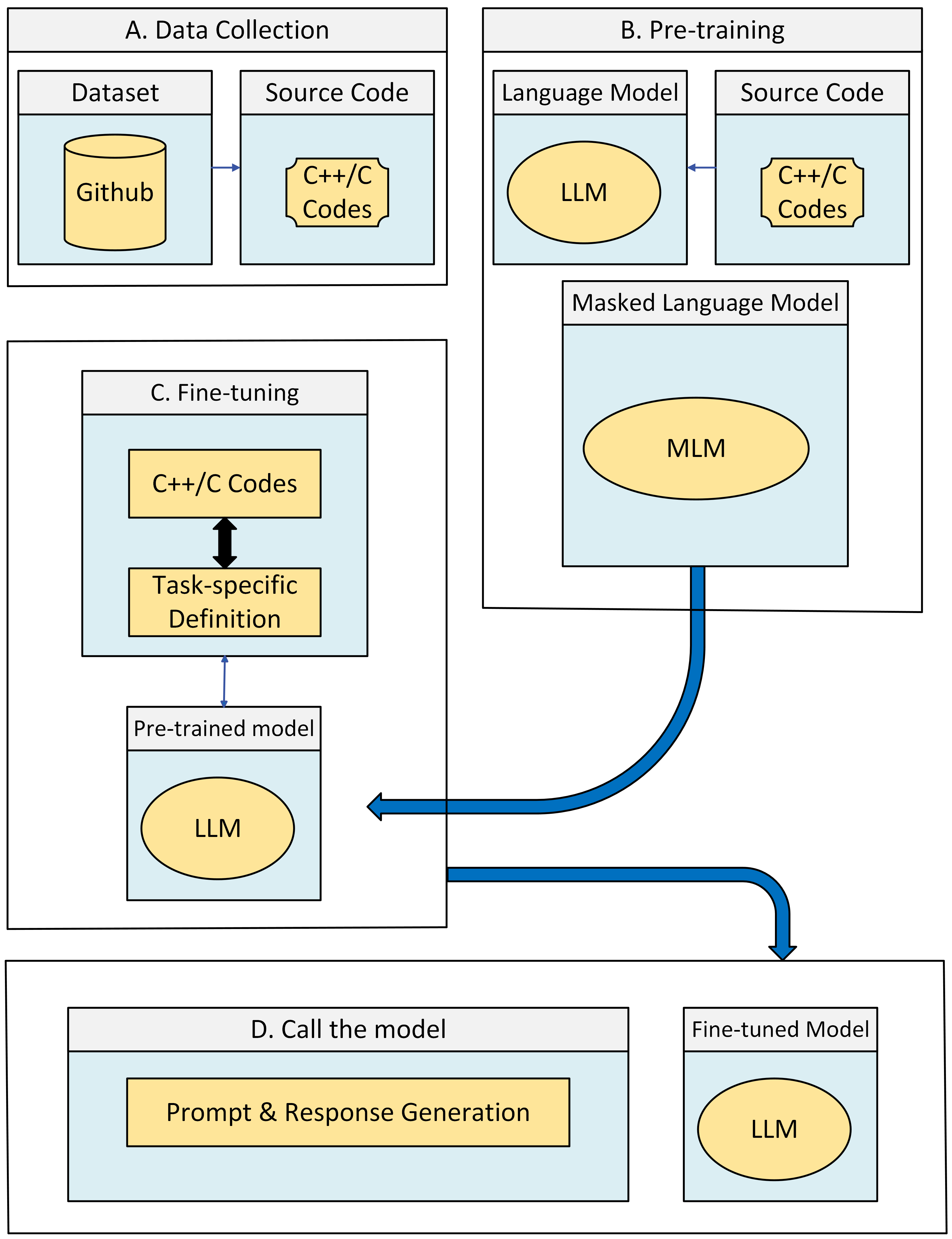}
    \caption{A general view of the pre-training of an LLM based on masked language modeling  }
  \label{fig:your-label}
\end{figure}

In \cite{159}, the authors proposed MonoCoder, a domain-specific language model tailored for high-performance computing (HPC) code and tasks. Pre-trained on HPCorpus—a dataset comprising C and C++ programs mined from GitHub—MonoCoder is focused exclusively on HPC-related code. In \cite{160}, the authors examined GitHub commits, which combine code changes with descriptive natural language messages to aid software developers in understanding software evolution. They pre-trained an LLM based on the BART model using 7.99 million commits across seven programming languages, resulting in a method called CommitBART. Similarly, in \cite{161}, the authors examined the effects of pre-training LLMs, such as GPT-2 (GPT2-Large) and TinyLlama v1.1, on programming languages versus natural languages for logical inference tasks. Decoder-based models were trained from scratch on datasets containing different programming languages, including Python, C, and Java, and other datasets (such as Wikipedia,) all under the same conditions.

In a different study, \cite{162} focused on pre-training and fine-tuning a GPT model with a large software codebase for automatic program repair (APR). They implemented a code-aware search strategy that prioritized compilable patches and solutions aligned with the length of the buggy code. Evaluations on the Defects4J \cite{just2014defects4j}  and QuixBugs \cite{lin2017quixbugs}  benchmarks demonstrated that CURE outperformed existing APR methods, fixing 57 and 26 bugs, respectively.\\

In \cite{163}, the authors introduced TreeBERT, a tree-based model aimed at improving tasks like code summarization and documentation. TreeBERT was pre-trained on datasets with different programming languages, using a hybrid objective that combined Tree Masked Language Modeling (TMLM) with Node Order Prediction (NOP). Additionally, in \cite{164}, the authors presented a new pre-training objective called Naturalizing for source code models built on CodeT5. Fine-tuned for tasks such as code generation, translation, and refinement, this model achieved state-of-the-art performance, particularly in zero-shot and few-shot learning tasks.\\

Furthermore, in \cite{165}, the authors evaluated the performance of pre-trained transformer models in software engineering tasks, including code summarization, bug detection, and understanding developer intent. They pre-trained BERT and SBERT on various datasets such as StackOverflow, GitHub issues, and Jira issues. Comparing BERT models (base and large) trained on software engineering data (e.g., code corpora) with those trained on general-domain data (e.g., Wikipedia). Table 
\ref{tab:PredictionMASK} shows the performance of different BERT models for [Mask] prediction. The study found that domain-specific pre-training yielded better results for tasks like bug prediction and code summarization. 

Generally, domain-specific pre-training for LLMs has emerged as a critical strategy for achieving superior performance in specialized tasks. By tailoring models to understand the unique nuances of a domain, researchers can unlock efficiencies and capabilities that general-purpose models may not achieve. These advancements not only enhance task-specific accuracy but also open up opportunities for groundbreaking applications, particularly in areas where domain expertise is paramount.

\begin{figure*}[p]  
\centering
\includegraphics[width=\textwidth,height=\textheight,keepaspectratio]{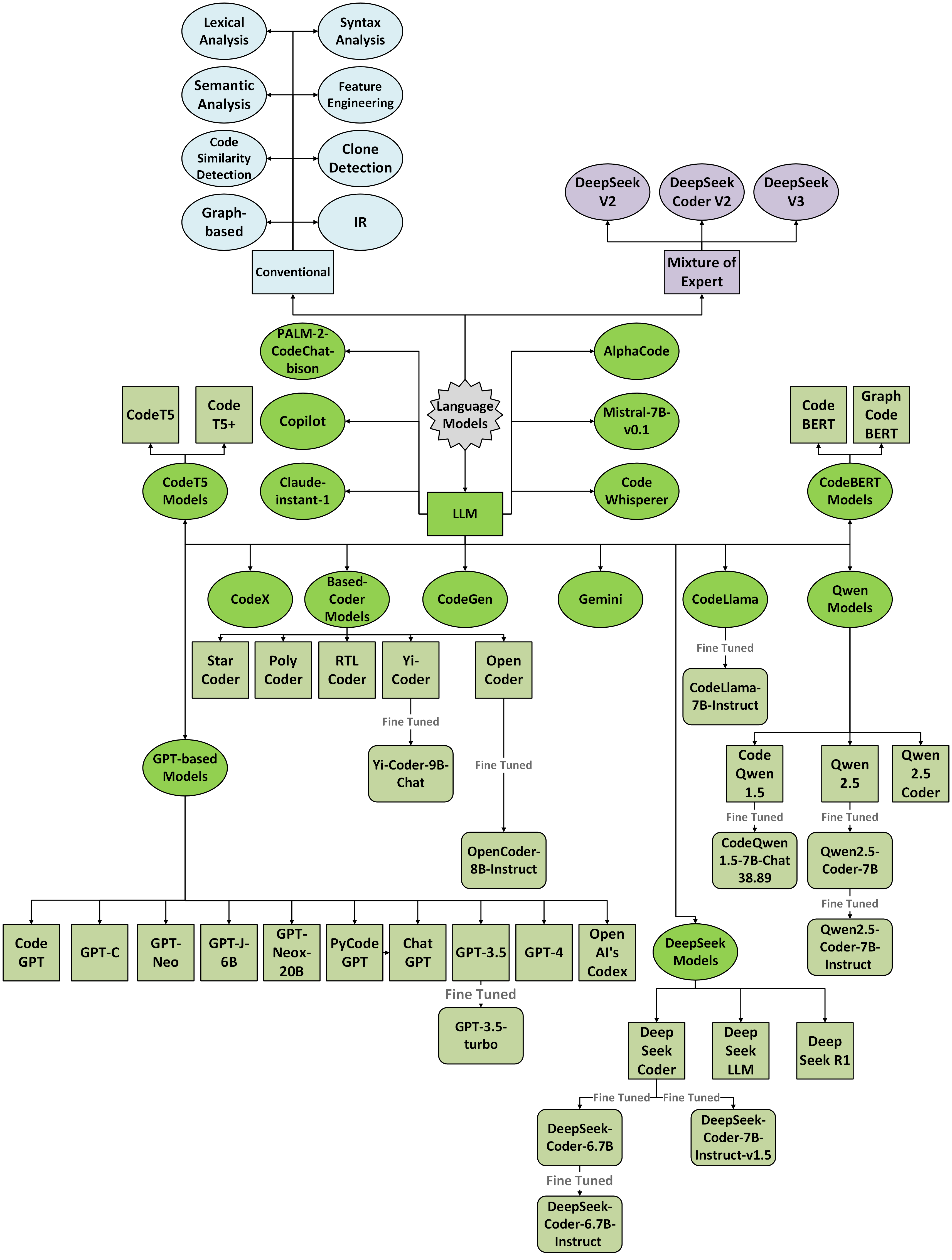}  
\caption{A taxonomy of recent NLP and LLM models for source code analysis}
\label{fig:Models-for-code-taxonomy}
\end{figure*}
\clearpage 

\section{Popular models: language models for source code analysis}
\begin{figure}[H]
    \centering
    \includegraphics[height = 11cm]{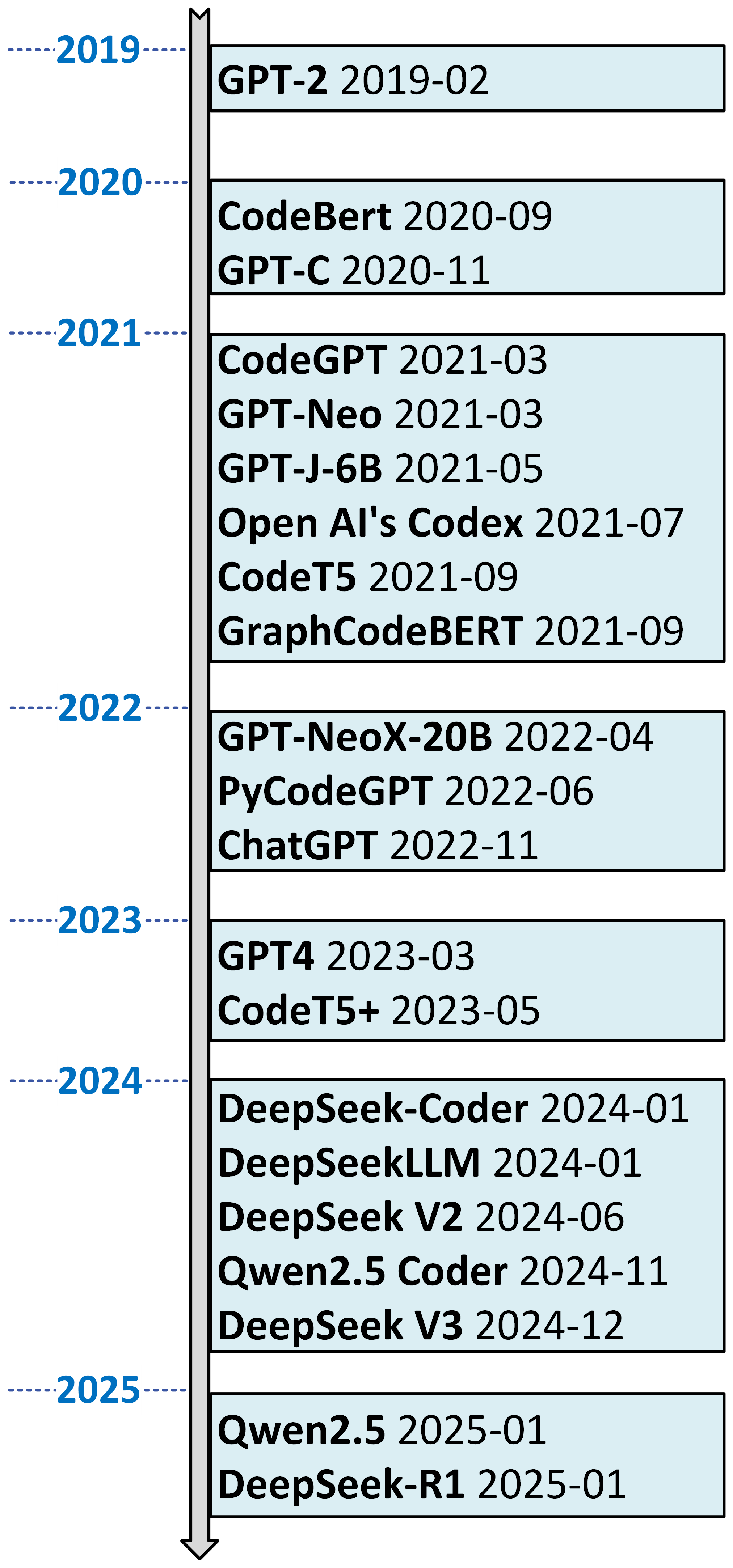}
    \caption{LLM model's timeline for source code analysis}
    \label{fig:LLM-Models-Timeline}
\end{figure}

\label{sec:Popular Models: Language models for Source
Code Analysis}
Although there are many LLM models for source code analysis, in this Section, we will focus only on the most famous models based on previous works which include CodeBERT, CodeT5, GPT, and DeepSeek models. In Figure \ref{fig:Models-for-code-taxonomy}, a taxonomy of language models for code is shown. Also, in figure \ref{fig:LLM-Models-Timeline} we bring the most important LLMs based on their technical report date.

\subsection{CodeBERT}
CodeBERT is a pre-trained bimodal transformer model for programming language (PL) and  natural language (NL) tasks, such as documentation generation and code search. It outperforms models like RoBERTa by combining replaced token detection (RTD) and masked language modeling (MLM) during training. CodeBERT uses both NL-PL pairs and unimodal data, boosting performance \cite{114}.

\subsection{CodeT5 Models}
~~~~CodeT5-based model are specially designed for code-related tasks \cite{113}. These models include CodeT5 and CodeT5+.

CodeT5 is a Transformer-based model, built on T5 architecture, which is applicable to both generating and understanding the code. CodeT5 employs multi-task learning enabling it to perform a variety of code-related tasks such as flaw detection, code translation and summarization. CodeT5 introduces identifier-aware pre-training objectives using crucial structural and semantic information in programming languages (PL). CodeT5 outperforms models like CodeBERT, PLBART, and GraphCodeBERT in multiple metrics across code-related tasks \cite{113}.

CodeT5+ is an extension to CodeT5, which is more flexible in architecture. Unlike CodeT5, it uses 3 architectures including encoder-only, decoder-only and encode-decoder. \cite{10}.
\subsection{Generative Pre-trained Transformer (GPT) Models}
The family of GPT models encompass different implementations including GPT-C \cite{176}, CodeGPT \cite{157}, GPT-Neo \cite{177}, GPT-J-6B \cite{178}, GPT-NeoX-20B \cite{179}, PyCodeGPT \cite{180}, ChatGPT \cite{181}, GPT-4 \cite{119} and OpenAI's Codex. Here we explain OpenAI's Codex \cite{106} and GPT-4 in more detail.

OpenAI's Codex is an open-source, GPT model written in Python which is evaluated on the HumanEval dataset, outperforming models like GPT-3 and GPT-J in functional correctness, solving 28.8\% of tasks with a single sample and up to 70.2\% with repeated sampling strategies to overcome complex prompts \cite{106}.

GPT 4 is a large-scale multi-modal model processing both text and image inputs and returning text outputs. It competes human-level performance on numerous tasks. The model achieves a better performance compared to previous versions, particularly on diverse language and reasoning tasks. GPT-4's high-level capabilities are achieved with advancements in pre-training and fine-tuning using reinforcement learning from human feedback (RLHF) \cite{119}.
\subsection{DeepSeek Models}
There are number of models implemented for DeepSeek family. We introduce the most important ones including DeepSeek-Coder \cite{116}, DeepSeek LLM \cite{117}, DeepSeek-V3 \cite{118} and DeepSeek-R1 \cite{168}.

DeepSeek-Coder is a series of open-source language models, ranging from 1.3B to 33B parameters, optimized for code intelligence in software development. Supporting 87 programming languages and using a 16K token context window, these models excel at tasks like code generation and completion, outperforming GPT 3.5 and Codex \cite{116,117}.

DeepSeek-V3, a Mixture-of-Experts (MoE) model with 671 billion parameters, uses Multi-head Latent Attention (MLA). It rivals top models like GPT-4 and Claude 3.5, with enhanced performance on Chinese tasks through Supervised Fine-Tuning (SFT) and Reinforcement Learning (RL). Its advanced optimization sets a new benchmark for open-source models, offering high performance with resource efficiency \cite{118}.

\subsection{Qwen Models}
There are various types of models in Qwen family, Qwen2.5-Coder \cite{213} and Qwen2.5 \cite{214} are the most important models.

The Qwen2.5-Coder is a series of LLMs developed by Alibaba's Qwen Team which is available in six sizes of 0.5B, 1.5B, 3B, 7B, 14B, and 32B parameters. These models are trained on 5.5 trillion tokens, with a focus on different code related tasks such as code generation, completion, debugging, and reasoning. They employ file-level and repository-level pretraining, and advanced strategies for fine-tuning. The Qwen2.5-Coder-32B model achieves the best results across HumanEval, MBPP and so on, outperforming open-source models like DeepSeek-Coder, and even competes with strong models like GPT-4o. The series is open-source and designed to facilitate real-world software development and automated programming \cite{213}.

Qwen2.5 is also developed by the same developer as Qwen2.5-coder, including mixture-of-Experts (MoE) models in sizes of 0.5B, 1.5B, 3B, 7B, 14B, 32B, and 72B parameters. It significantly improves over previous versions with pre-training on 18 trillion tokens, expanded instruction fine-tuning, and reinforcement learning for enhanced customized personalization. The series supports long-context processing (up to 1M tokens in Qwen2.5-Turbo) and optimized for reasoning, coding, mathematics, and multilingual tasks. Qwen2.5 outperforms major open-source models like Llama-3-405B while maintaining a strong cost-performance trade-off. \cite{214}.
\clearpage
\section{Datasets used for LLM code analysis}
\label{sec:Datasets and LLM code analysis}
Using public datasets for LLM training and source code analysis is essential for advancing the capabilities of large language models in programming-related tasks. Here, we provide a list of datasets that can be utilized for various LLM tasks, such as code summarization and and code analysis. 

 
Table \ref{tab:Public and free datasets for source code analysis tasks} shows a list of free datasets that we can we use for various LLM tasks for source code analysis. For example. CodeXGLUE is a widely-used benchmark that includes datasets for tasks like code summarization, translation, and completion \cite{157}. It supports multiple programming languages, including Python, Java, and JavaScript. The data is often provided in formats like JSON or CSV, with fields for code, documentation, and input-output pairs tailored for fine-tuning.

Figure \ref{fig:LLM-Datasets-Timeline} show a time-line of free datasets for code analysis, we can mention famous datasets such as CodesearchNet, CodexGLU, TheStack, and CodeNet, which are key in code search and programming language research. CodesearchNet enables code retrieval through natural language queries, covering multiple programming languages. CodexGLUE focuses on improving code generation and completion via natural language processing. TheStack offers a large collection of code snippets across various languages, supporting cross-lingual code search.\\\\

Another notable dataset is CodeNet\cite{157}, is a large-scale collection containing around 14 million code samples, each representing a solution to one of nearly 4,000 coding problems. The dataset includes code written in over 50 programming languages, with a particular focus on C++, C, Python, and Java. This dataset, developed by IBM, provides a vast collection of labeled code for program understanding and classification tasks. These datasets are crucial for developing LLM models for code analysis and intelligent code search. In figure \ref{fig:LLM-Datasets-Timeline} the initial release date for each data set is shown:
\begin{figure}[H]
    \centering
    \includegraphics[width=0.6\linewidth]{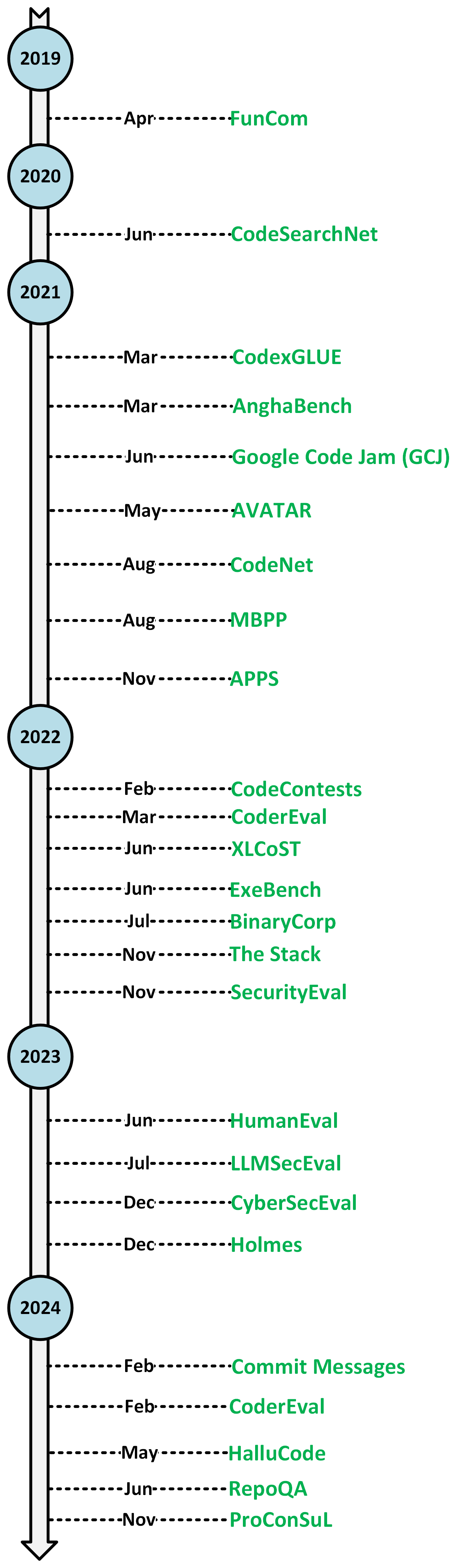}
    \caption{Time Line of LLM Datasets.}
    \label{fig:LLM-Datasets-Timeline}
\end{figure}



\clearpage
\begin{table*}[p]
\centering
\caption{Publicly available datasets for source code analysis tasks.}
\resizebox{\textwidth}{!}{ 

\begin{tabular}{l l l l}
\hline
Datasets                       & Details/Langauages                                                                                                                                                                                                                                                                                                                                                                                                  &  Tasks                                                                 & Link                                                          \\ \hline
\multirow{3}{*}{CodeSearchNet} & \multirow{2}{*}{\begin{tabular}[c]{@{}l@{}}2M (comment, code) pairs\end{tabular}}                                                                                                                                                                                                                                                & \multirow{3}{*}{\begin{tabular}[c]{@{}l@{}}Fine-tunning/ \\ Pre-training\end{tabular}} & \multirow{3}{*}{https://github.com/github/CodeSearchNet}      \\
                               &                                                                                                                                                                                                                                                                                                                                                                                                                     &                                                                                        &                                                               \\ 
                               & \begin{tabular}[c]{@{}l@{}}\\ Ruby\end{tabular}                                                                                                                                                                                                                                                                                                                                  &                                                                                        &                                                               \\ \hline
CommitMessages                 & CommitMessages                                                                                                                                                                                                                                                                                                                                                                                                      & Pre-training                                                                           & https://github.com/src-d/datasets/tree/master/CommitMessages  \\ \hline
CodeNet                        & \begin{tabular}[c]{@{}l@{}}CodeNeta total of\\13,916,868 submissions,\\  divided into \\4053 problems\\/ C/C++\end{tabular}                                                                                                                                                                                                                                                                                           & Fine-tunning                                                                           & https://github.com/IBM/Project\_CodeNet                       \\ \hline
XLCoST Dataset                 & \begin{tabular}[c]{@{}l@{}}for code generation, \\ code translation\\(code-to-code), \\ code summarization\\(code-to-text), \\ and code synthesis\\(text-to-code).\end{tabular}                                                                                                                                                                                                                                        & Fine-tunning                                                                           & https://github.com/reddy-lab-code-research/XLCoST             \\ \hline
The stack                      & \begin{tabular}[c]{@{}l@{}}6TB of code,\\ 358 languages.\end{tabular}                                                                                                                                                                                                                                                            &                                                                                        & https://huggingface.co/datasets/bigcode/the-stack             \\ \hline
AVATAR                         & \begin{tabular}[c]{@{}l@{}}A collection of\\9515 programming \\ problems and their\\solutions written\\ ( Java and Python)\end{tabular}                                                                                                                                                                                                                                                                              & \begin{tabular}[c]{@{}l@{}}Function\\translation\end{tabular}                                                                  & https://github.com/wasiahmad/AVATAR                           \\ \hline
MBPP                           & \begin{tabular}[c]{@{}l@{}}1,000 crowd-sourced \\ Python problems\end{tabular}                                                                                                                                                                                                                                                                                                                                      & Pre-training                                                                           & https://huggingface.co/datasets/google-research-datasets/mbpp \\ \hline
ProConSuL                      & \begin{tabular}[c]{@{}l@{}}GitHub repositories\\written in C/C++  \end{tabular}                                                                                                                                                                                                                                                                                                                                                                              & \begin{tabular}[c]{@{}l@{}}SFT dataset/\\ Code\\summarization\end{tabular}              & https: //github.com/trinity4ai/ProConSuL.                   \\ \hline
\multirow{2}{*}{CodeXGLUE}     & \begin{tabular}[c]{@{}l@{}}A benchmark for program \\ understanding\end{tabular}                                                                                                                                                                                                                                                                                                                                    & \multirow{2}{*}{General}                                                               & \multirow{2}{*}{https://github.com/microsoft/CodeXGLUE}       \\ 
                               & \begin{tabular}[c]{@{}l@{}}14 datasets \\for different tasks:\\code-code, text-code,\\code-text, and text-text)\end{tabular} &                                                                                        &                                                               \\ \hline
\end{tabular}
\label{tab:Public and free datasets for source code analysis tasks}
}
\end{table*}

\begin{table*}[p]
    \centering
    \caption{Prediction of words for [MASK] Tokens by pre-training on BERT models (A snapshot taken from \cite{165}) }
    \resizebox*{\textwidth}{!}{
    \begin{tabular}{l l l l l l}

        \hline

        \begin{tabular}[c]{@{}l@{}}1) Pathlib is a python library\\used for handling [MASK].\end{tabular} &\begin{tabular}[c]{@{}l@{}}paths\end{tabular}&\begin{tabular}[c]{@{}l@{}}applications\\systems\\programs\\software\\data\end{tabular}&\begin{tabular}[c]{@{}l@{}}applications\\software\\programs\\languages\\systems\end{tabular}&\begin{tabular}[c]{@{}l@{}}paths\\files\\urls\\URLs\\pathnames\end{tabular}&\begin{tabular}[c]{@{}l@{}}paths\\path\\directories\\filenames\\strings\end{tabular}\\

        \hline

        \begin{tabular}[c]{@{}l@{}}2) I have to discuss this\\with the other [MASK].\end{tabular} &\begin{tabular}[c]{@{}l@{}}developers\end{tabular}&\begin{tabular}[c]{@{}l@{}}elders\\girls\\men\\officers\\council\end{tabular}&\begin{tabular}[c]{@{}l@{}}men\\members\\elders\\officers\\people\end{tabular}&\begin{tabular}[c]{@{}l@{}}guys\\people\\developers\\person\\users\end{tabular}&\begin{tabular}[c]{@{}l@{}}developers\\team\\maintainers\\people\\devs\end{tabular}\\
        \hline
        
    \end{tabular}
    }
    \label{tab:PredictionMASK}
\end{table*}
\clearpage

\section{Discussion and Future work}
\label{sec:Discussion and Future works}
Regarding the application of LLMs for source code analysis presents several challenges and opportunities for future research. Based on our studies, there are key areas that demand further exploration, which hold significant potential for advancing this field. In this context, we identify seven critical issues and observe that the following gaps remain insufficiently addressed. These gaps highlight promising directions for future work in source code analysis using LLMs.

\subsection{Limitations of Publicly Available Datasets for LLM Tasks} 
Although there are existing datasets for source code such as CodeSearchNet, The Stack or Commit  Messages , there remain significant limitations when it comes to finding specialized datasets tailored for specific tasks related to fine-tuning LLMs for source code analysis. Tasks such as code summarization, code disassembly, code decompiling, and comment generation require datasets that focus on the nuances of these specific activities.\\

While general-purpose code datasets can provide a foundation for training LLMs on basic code generation tasks, they often do not address the complexities and requirements of these advanced tasks \cite{166,167}. Another important aspect to consider is the quality of source code. For instance, in \cite{169}, the authors encountered challenges with the quality of the assembly code search dataset, as it depended on docstrings from sources like CodeSearchNet. In CodeSearchNet's evaluation, 32.8\% of docstrings were irrelevant to the source code.
Furthermore, comment generation involves aligning natural language with specific code functionalities, a task that is not always explicitly covered by standard code datasets. Our survey reveals a notable gap in the availability of specialized datasets for fine-tuning LLMs on specific tasks, including code summarization, disassembly, decompilation, and comment generation. While existing code datasets can support basic code generation tasks, they fail to address the complexities involved in these advanced code analysis tasks. This limitation underscores the need for the creation of more targeted datasets that focus on these specific areas, enabling LLMs to be fine-tuned for tasks that require deeper understanding and nuanced interactions with source code. Developing such datasets would be a critical step toward improving the capabilities of LLMs in real-world software analysis and development environments.

\subsection{Long Code analysis and Token Size} 
Working with long code in a language model requires careful handling to ensure that the model doesn’t exceed its token limit and that context is preserved across different chunks of the code. For example, some models like GPT-4 have a maximum token limit \cite{172,173,174,175} , which includes both the prompt and the response. This means that when working with large codebases, only a portion of the code may be processed at a time, potentially leading to truncation or loss of important context. As a result, it’s crucial to find ways to manage large code inputs effectively.\\

For example in \cite{170} , the authors encountered challenges and limitations due to the token size constraints of transformer-based models, impacting the generation of CodeBERT embeddings. To fit within the 512-token limit, we had to truncate the tokens, potentially leading to the loss of some syntactic information from the code snippets.

To address this challenge, it would be helpful to consider methods for chunking the code into smaller sections while maintaining logical consistency \cite{189}. For instance, breaking down code into functions, classes, or modules and analyzing them incrementally can help reduce the token count per request while allowing the model to process the code in smaller, more manageable parts. However, careful attention must be given to ensure that context is not lost between chunks, which could result in incorrect analysis or suggestions\cite{190,191}. This is particularly critical for codebases that are interdependent, where variables or functions defined earlier may be referenced later.\\

Moreover, hybrid approaches combining language models with other techniques can be effective for handling large-scale code analysis. Non-LLM methods \cite{192}, such as static code analyzers, or even specialized machine learning models trained for code tasks, could be used in collaboration with LLMs to perform complementary tasks like detecting syntax errors, identifying security vulnerabilities, or suggesting refactoring improvements. These methods can help alleviate some of the limitations of LLMs, particularly when it comes to scalability and accuracy in analyzing long code.\\

Additionally, newer models or extensions, such as those designed to handle longer token sequences like Longformer \cite{beltagy2020longformer} or Reformer, could be explored to overcome token limitations inherent in traditional transformer-based models \cite{193}. These models are designed to process longer input sequences more efficiently by using techniques like sparse attention mechanisms, which allow them to handle larger contexts without overwhelming the model’s capacity.

\subsubsection{ Long code analysis and Prompt chaining}
Prompt chaining is a technique used to break down complex tasks into smaller, sequential prompts, enabling efficient processing of large-scale data while overcoming token limitations. A related study worth mentioning is LLM4FL \cite{171}, where the authors tackled token limitations and complexity in fault localization. Their approach utilizes a divide-and-conquer strategy, incorporating prompt chaining and multiple LLM agents to enhance efficiency. It splits large-scale coverage data into manageable chunks, allowing each LLM agent to analyze a subset of the code independently. By integrating Spectrum-Based Fault Localization (SBFL) rankings, LLM4FL refines fault detection across multiple iterations, cross-referencing results between agents to improve accuracy. This structured approach ensures efficient fault localization in complex systems while overcoming performance degradation with long inputs.
\subsection{Most Used Models: DeepSeek vs. GPT-4 Family}
\label{sec:DeepSeek Vs. GPT4}
In our review, most of the work focused on two main families of models: Deepseek models, such as \cite{10,13,22,26,52,168}, and GPT models, such as \cite{11,15,19,23,28}. In table \ref{tab:DeepSeekR1-Vs-GPT4} we make a comprehensive comparison between these two models.
\begin{table}[h]
    \centering
    \caption{DeepSeek-R1 vs GPT-4: strong (S), super strong (SS), super super strong (SSS), software engineering (SE), reinforcement learning from human feedback (RLHF)}
    \resizebox{0.5\textwidth}{!}{
    \begin{tabular}{l l l}
        \hline
        Criteria & GPT-4 & DeepSeek-R1 \\
        \hline
        Input Type & text, multimedia & text\\
        \hline
        Developer & OpenAI & DeepSeek-AI\\
        \hline
        \begin{tabular}[c]{@{}l@{}}Training\\Approach\end{tabular} & \begin{tabular}[c]{@{}l@{}}Transformation-based\\Pre-Training + RLHF\end{tabular} & RL + Cold Start\\
        \hline
        \makecell{Training Data} & Public + Licensed & Public + Qwen + Llama\\
        \hline
        Architecture & Transformer & Transformer\\
        \hline
        Fine Tuning & RLHF & RL + Distillation\\
        \hline
        \begin{tabular}[c]{@{}l@{}}Reasoning\\Performance\end{tabular} & S & SS\\
        \hline
        \begin{tabular}[c]{@{}l@{}}Mathematical\\Ability\end{tabular} & S & SSS\\
        \hline
        \begin{tabular}[c]{@{}l@{}}Code\\Generation\end{tabular} & S & SSS\\
        \hline
        \begin{tabular}[c]{@{}l@{}}General\\Knowledge\end{tabular} & S & S\\
        \hline
        \begin{tabular}[c]{@{}l@{}}Context\\Length\end{tabular} & Limited & Longer\\
        \hline
        \begin{tabular}[c]{@{}l@{}}Safety\\Mechanism\end{tabular} & Anti Adversary & Avoids Bias\\
        \hline
        \begin{tabular}[c]{@{}l@{}}Is Open\\Source?\end{tabular} & No & Yes\\
        \hline
        \begin{tabular}[c]{@{}l@{}}Special\\Ability\end{tabular} & \begin{tabular}[c]{@{}l@{}}General Purpose,\\Supporting Multi-media\end{tabular} & \begin{tabular}[c]{@{}l@{}}High Performance\\for Reasoning\end{tabular}\\
        \hline
        limitations & \begin{tabular}[c]{@{}l@{}}Not Fully Reliable\\Biased Answers\end{tabular} & \begin{tabular}[c]{@{}l@{}}Excelling only in\\English and Chinese,\\Prompt Sensitivity,\\Limited Capability\\in SE Tasks\end{tabular}\\
        \hline
        \makecell{Release Date} & 2023-03 & 2025-01\\
        \hline
        Cost & High & Low\\
        \hline
        \begin{tabular}[c]{@{}l@{}}Resource\\Consumption\end{tabular} & High & Low\\
        \hline
    \end{tabular}
    }
    \label{tab:DeepSeekR1-Vs-GPT4}
\end{table}

Regarding the table \ref{tab:DeepSeekR1-Vs-GPT4}, while both models use transformer architectures and use public datasets, DeepSeek-R1 integrates additional data from Qwen and Llama, along with distillation techniques for fine-tuning. Performance-wise, DeepSeek-R1 surpasses GPT-4 in reasoning, mathematical ability, and code generation, demonstrating "super super strong" (SSS) capabilities in these areas. Additionally, DeepSeek-R1 features a longer context length and an open-source framework, whereas GPT-4 remains proprietary and employs anti-adversarial safety mechanisms.\\

However, in summary, both DeepSeek and GPT models are powerful language models for source code analysis tasks. The choice between these models can depend on the specific targets and tasks, as each may offer distinct advantages depending on the nature of the problem being addressed \cite{116,186,187}. For example, DeepSeek is an open-source model that excels in tasks requiring detailed semantic understanding and structured code analysis. On the other hand, GPT models \cite{188}, known for their ability to generate human-like text, are highly versatile and can handle a wide range of tasks, from code completion to generating documentation and explanations for complex code segments.

\subsection{Quality of LLMs for code summarization task}

Recent findings have evaluated the effectiveness of LLMs for code summarization tasks \cite{28,30,33,35,39,40,44}. While certain aspects of code summarization are gaining attention in the era of LLMs, but they are still affected by some challenges. One of the major limitations of LLMs for code summarization is their inability to validate code correctness or test its runtime behavior \cite{54,181}. While they can generate summaries based on code syntax, they cannot run the code to check for issues like bugs or edge cases\cite{183}. In some works in \cite{184}, the authors studied LLM-based code summarization, compared automated evaluation methods (including GPT-4) with human assessments, and found that GPT-4 had the strongest correlation. They also tested five prompting techniques for Java, Python, and C, discovering that simpler zero-shot prompting could outperform more advanced methods depending on the LLM and language.

\section{Conclusion}
\label{sec:Conclusion}
NLP and LLMs are transforming code analysis in computer science by understanding both the structure and intent behind code. In this study, we explored the applications of LLMs for the code analysis, focusing on their ability to assist in tasks such as code summarization, code generation, comment generation, and disassembling, and decompiling code. We also investigated datasets and famous LLM models for the source code analysis. We believe this research opens up new opportunities for using LLMs in software development, offering valuable insights for both developers and researchers in the field. However, based on our understanding of the surveyed topic, we summarize the major lessons we learned as follows:

\begin{itemize}
  \item High-quality datasets and model selection are crucial for effective source code analysis. The accuracy and performance of LLMs depend significantly on the training data used, as models learn to recognize patterns and generate meaningful outputs based on the data they have been exposed to. Datasets such as CodeXGLUE, CodeNet, and GitHub repositories provide diverse code examples that help train LLMs to understand different programming languages, styles, and structures. However, dataset quality varies, and biases present in training data can impact model reliability. Additionally, selecting the right LLM model for a specific task is essential. OpenAI Codex, Code Llama, and AlphaCode each offer different strengths, and their performance varies depending on the complexity of the code analysis required. Choosing the most suitable model and training it on high-quality, well-labeled datasets can significantly enhance the effectiveness of AI-driven code analysis.
  
  \item Interdisciplinary research is key to advancing LLM-based code analysis. This field lies at the intersection of natural language processing, machine learning, and software engineering, and progress in any of these areas directly impacts the performance of LLMs in code analysis. Collaboration between researchers from different domains can lead to innovative approaches, such as integrating formal methods with machine learning to enhance the accuracy of automated code verification. Additionally, combining techniques from job scheduling, optimized data storage, and efficient code representation can help improve the scalability and effectiveness of LLM-driven tools. By fostering interdisciplinary collaboration, researchers can develop more advanced AI-powered systems that understand and process source code more effectively.
  
  \item Customization and fine-tuning of LLMs enhance their performance for specific programming domains. While general-purpose LLMs demonstrate impressive capabilities, they may not always provide optimal results for specialized fields such as embedded systems, cybersecurity, or financial software. Training or fine-tuning models on domain-specific datasets can significantly improve their accuracy and relevance for targeted applications. For instance, an LLM trained on financial transaction processing code may better understand compliance requirements and security constraints compared to a general-purpose model. Customization allows developers to tailor AI-driven tools to their unique needs, leading to more precise and context-aware code analysis.

\end{itemize}

We are standing at the forefront of LLM-powered source code analysis, and its integration into software engineering presents both exciting opportunities and significant challenges. As LLMs continue to evolve, their ability to automate and enhance various aspects of coding will improve, but further research is needed to optimize their efficiency, accuracy, and security. We hope this study serves as a foundation for future research, inspiring further exploration into the capabilities and limitations of LLMs in source code analysis.


\bibliographystyle{unsrt}
\bibliography{main}

\end{document}